\documentclass[12pt, a4paper]{article}

\usepackage[a4paper, top=2.0cm, bottom=1.0cm, left=1.8cm, right=1.5cm, footskip = 1.0cm, includehead, includefoot]{geometry}
\usepackage{helvet} 			

\usepackage[utf8]{inputenc}
\usepackage{csquotes}

\usepackage[final]{graphicx}
\usepackage[margin=20mm, font=footnotesize, labelfont=bf, justification=justified]{caption}
\usepackage{dirtytalk}

\usepackage{graphicx}        	
\usepackage{booktabs} 
\usepackage{caption}                             		 	
\usepackage{indentfirst}
\usepackage{csquotes}

\usepackage{times}
\usepackage[bottom]{footmisc}	
\usepackage{scrextend}
\deffootnote[1.8em]{1.8em}{1.8em}{\textsuperscript{\thefootnotemark}}
\usepackage{xcolor}			
\usepackage{newtxtext}
\usepackage{newtxmath}       	
\usepackage{braket}
\usepackage{float}			
\usepackage{mathtools}

\DeclareUnicodeCharacter{2212}{-}

\usepackage{nicematrix}
\usepackage{amsmath}
\usepackage{subfigure}
\usepackage{tipa}
\usepackage{url}
\usepackage{authblk}
\usepackage{blindtext}
\usepackage{hyperref}
\usepackage{xcolor}
\usepackage[capitalize]{cleveref}

\usepackage{titlesec} 
\titleformat{\paragraph}
{\normalfont\normalsize\bfseries}{\theparagraph}{1em}{}

\usepackage{musicography}  

\usepackage{physics}
\usepackage{braket} 

\usepackage{musicography}  

\usepackage{tikz} 
\usetikzlibrary{quantikz2}
\usetikzlibrary{matrix}

\usepackage{MnSymbol} 
\usepackage{hyperref}
\usepackage{tikzfig}
\usepackage{relsize}
\usepackage[outline]{contour}
\contourlength{.12em}

\tikzstyle{node}=[minimum size=0.3cm]
\tikzstyle{Z}=[fill={rgb,255:red,230; green,254; blue,230}, draw={rgb,255: red,61; green,77; blue,61}, shape=circle]
\tikzstyle{X}=[fill={rgb,255:red,255; green,135; blue,136}, draw={rgb,255: red,102; green,54; blue,54}, shape=circle]
\tikzstyle{Y}=[fill=zxblue, draw=zxdblue, shape=circle]
\tikzstyle{Z_big}=[fill={rgb,255:red,230; green,254; blue,230}, draw={rgb,255: red,61; green,77; blue,61}, shape=circle, minimum width=1.6em, font={\small}]
\tikzstyle{X_big}=[fill={rgb,255:red,255; green,135; blue,136}, draw={rgb,255: red,102; green,54; blue,54}, shape=circle, minimum width=1.6em, font={\small}]
\tikzstyle{Y_big}=[fill=zxblue, draw=zxdblue, shape=circle, minimum width=1.6em, font={\small}]
\tikzstyle{Z_tri}=[fill={rgb,255:red,230; green,254; blue,230}, draw={rgb,255: red,61; green,77; blue,61}, regular polygon, regular polygon sides=3, draw, shape border rotate=0, inner sep=0pt, minimum width=15pt, line width=0.75]
\tikzstyle{X_tri}=[fill={rgb,255:red,255; green,135; blue,136}, draw={rgb,255: red,102; green,54; blue,54}, regular polygon, regular polygon sides=3, draw, shape border rotate=0, inner sep=0pt, minimum width=15pt, line width=0.75]
\tikzstyle{Z_tri_inv}=[fill={rgb,255:red,230; green,254; blue,230}, draw={rgb,255: red,61; green,77; blue,61}, regular polygon, regular polygon sides=3, draw, shape border rotate=180, inner sep=0pt, minimum width=15pt, line width=0.75, font={\footnotesize}]
\tikzstyle{X_tri_inv}=[fill={rgb,255:red,255; green,135; blue,136}, draw={rgb,255: red,102; green,54; blue,54}, regular polygon, regular polygon sides=3, draw, shape border rotate=180, inner sep=0pt, minimum width=15pt, line width=0.75]
\tikzstyle{Z_tri_l}=[fill={rgb,255:red,230; green,254; blue,230}, draw={rgb,255: red,61; green,77; blue,61}, regular polygon, regular polygon sides=3, draw, shape border rotate=90, inner sep=0pt, minimum width=15pt, line width=0.75]
\tikzstyle{X_tri_l}=[fill={rgb,255:red,255; green,135; blue,136}, draw={rgb,255: red,102; green,54; blue,54}, regular polygon, regular polygon sides=3, draw, shape border rotate=90, inner sep=0pt, minimum width=15pt, line width=0.75]
\tikzstyle{H}=[fill=yellow, draw=black, shape=rectangle, minimum width=2mm, minimum height=2mm]
\tikzstyle{Y_box}=[draw=black, shape=rectangle, minimum width=2mm, minimum height=2mm, tikzit fill={rgb,255: red,255; green,128; blue,0}]
\tikzstyle{Z_med}=[fill={rgb,255:red,230; green,254; blue,230}, draw={rgb,255: red,61; green,77; blue,61}, shape=circle, minimum width=1.1em, font={\footnotesize}]
\tikzstyle{X_med}=[fill={rgb,255:red,255; green,135; blue,136}, draw={rgb,255: red,102; green,54; blue,54}, shape=circle, minimum width=1.1em, font={\footnotesize}]
\tikzstyle{Y_med}=[fill=zxblue, draw=zxdblue, font={\footnotesize}, minimum width=1.1em, font={\footnotesize}, shape=circle]
\tikzstyle{ZP}=[fill={rgb,255:red,230; green,254; blue,230}, draw={rgb,255: red,61; green,77; blue,61}, regular polygon, regular polygon sides=3, shape border rotate=30]
\tikzstyle{ZM}=[fill={rgb,255:red,230; green,254; blue,230}, draw={rgb,255: red,61; green,77; blue,61}, regular polygon, regular polygon sides=3, shape border rotate=-30]
\tikzstyle{XP}=[fill={rgb,255:red,255; green,135; blue,136}, draw={rgb,255: red,102; green,54; blue,54}, regular polygon, regular polygon sides=3, shape border rotate=30]
\tikzstyle{XM}=[fill={rgb,255:red,255; green,135; blue,136}, draw={rgb,255: red,102; green,54; blue,54}, regular polygon, regular polygon sides=3, shape border rotate=-30]
\tikzstyle{small_box}=[fill=white, draw=black, shape=rectangle, minimum width=1.5cm, minimum height=1.5cm, font={\footnotesize}]
\tikzstyle{med_rectangle}=[fill=white, draw=black, shape=rectangle, minimum width=2.8cm, minimum height=1.5cm, font={\footnotesize}]
\tikzstyle{big_rectangle}=[fill=white, draw=black, shape=rectangle, minimum width=4.5cm, minimum height=1.5cm, font={\footnotesize}]
\tikzstyle{smol_box}=[fill=white, draw=black, shape=rectangle, minimum width=1cm, minimum height=1cm]
\tikzstyle{poo}=[minimum height=0.7cm, minimum width=0.7cm, path picture={\node at (path picture bounding box.center) {\includegraphics[width=0.7cm] {figures/poo}};}]
\tikzstyle{Z_long}=[fill={rgb,255:red,230; green,254; blue,230}, draw={rgb,255: red,61; green,77; blue,61}, shape=rectangle, rounded corners=0.25cm, minimum height=0.5cm, inner sep=0.25em, font={\scriptsize}]
\tikzstyle{X_long}=[fill={rgb,255:red,255; green,135; blue,136}, draw={rgb,255: red,102; green,54; blue,54}, shape=rectangle, rounded corners=0.25cm, minimum height=0.5cm, inner sep=0.25em, font={\scriptsize}]
\tikzstyle{med_rectv}=[fill=white, draw=black, shape=rectangle, minimum width=1.1cm, minimum height=2.4cm, font={\scriptsize}, inner sep=0.2cm]
\tikzstyle{Z dot}=[fill={rgb,255: red,0; green,127; blue,0}, draw=black, shape=circle, minimum width=1.5mm]
\tikzstyle{X dot}=[fill={rgb,255:red,255; green,21; blue,0}, draw=black, shape=circle, minimum width=1.5mm]
\tikzstyle{Z phase dot}=[fill={rgb,255: red,0; green,127; blue,0}, draw=black, shape=circle, font={\footnotesize}]
\tikzstyle{X phase dot}=[fill={rgb,255:red,255; green,21; blue,0}, draw=black, shape=circle, font={\footnotesize}]
\tikzstyle{ZYa}=[draw=black, shape=rectangle, rectangle split, rectangle split parts=2, rectangle split horizontal, rectangle split part fill={zxgreen, zxblue}, rectangle split draw splits=false, minimum height=2mm, font={\tiny}]
\tikzstyle{YZ}=[draw=black, shape=rectangle, rectangle split, rectangle split parts=2, rectangle split horizontal, rectangle split part fill={zxblue, zxgreen}, rectangle split draw splits=false, minimum height=2mm, font={\tiny}]
\tikzstyle{XYa}=[draw=black, shape=rectangle, rectangle split, rectangle split parts=2, rectangle split horizontal, rectangle split part fill={zxred, zxblue}, rectangle split draw splits=false, minimum height=2mm, font={\tiny}]
\tikzstyle{YX}=[draw=black, shape=rectangle, rectangle split, rectangle split parts=2, rectangle split horizontal, rectangle split part fill={zxblue, zxred}, rectangle split draw splits=false, minimum height=2mm, font={\tiny}]
\tikzstyle{tiny_box}=[fill=white, draw=black, shape=rectangle, minimum width=1cm, minimum height=1cm, font={\footnotesize}]
\tikzstyle{scalar}=[fill=white, draw=black, shape=diamond, font={\scriptsize}]
\tikzstyle{black_dot}=[fill=black, draw=black, shape=circle, inner sep=0pt, minimum size=0.2cm, text height=2pt, text depth=0pt]
\tikzstyle{white_dot}=[fill=none, draw=black, shape=circle, inner sep=0pt, minimum size=0.4cm]

\tikzstyle{dashs}=[-, dashed, line width=0.15mm]
\tikzstyle{thick}=[-, line width=0.5mm]
\tikzstyle{arrow}=[->]
\tikzstyle{invisible}=[-, draw=none]
\tikzstyle{functor}=[-, fill={rgb,255: red,240; green,240; blue,240}]
\tikzstyle{boxedge}=[-, fill=white]
\tikzstyle{dashed_line}=[dashed, gray, line width=0.25mm]

\tikzstyle{black_dot}=[fill=black, draw=black, shape=circle, inner sep=0pt, minimum size=0.1cm, text height=2pt, text depth=0pt]
\tikzstyle{white_dot}=[fill=none, draw=black, shape=circle, inner sep=0pt, minimum size=0.2cm]

\tikzstyle{dashed_line}=[dashed, gray, line width=0.25mm]

\tikzstyle{color1}=[fill={rgb,255: red,155; green,155; blue,155}, line width=0.25mm]
\tikzstyle{color2}=[fill={rgb,255: red,180; green,180; blue,180}, line width=0.25mm]
\tikzstyle{color3}=[fill={rgb,255: red,205; green,205; blue,205}, line width=0.25mm]
\tikzstyle{color4}=[fill={rgb,255: red,230; green,230; blue,230}, line width=0.25mm]
\tikzstyle{color5}=[fill={rgb,255: red,255; green,255; blue,255}, line width=0.25mm]

\usepackage{listings}

\makeatletter
\def\mathcolor#1#{\@mathcolor{#1}}
\def\@mathcolor#1#2#3{%
  \protect\leavevmode
  \begingroup
    \color#1{#2}#3%
  \endgroup
}
\makeatother

\definecolor{light-gray}{gray}{0.95}
\newcommand{\code}[1]{\colorbox{light-gray}{\texttt{#1}}}

\title{\huge Teleportation Game: Quantum Teleportation in Multi-Agent Systems for Interactive Music}
\date{}
\author[1]{Eduardo Reck Miranda}
\author[2]{Scott Yeiichi Oshiro}

\affil[1]{University of Plymouth, Interdisciplinary Centre for Computer Music Research (CCMR), Plymouth, UK, eduardo.miranda@plymouth.ac.uk}
\affil[2]{Stanford University, Anesthesiology, Perioperative and Pain Medicine, Stanford, CA, USA, soshiro1@stanford.edu }

\begin{document}

\maketitle

\abstract{This paper introduces an interactive music system with quantum musical agents that communicate by teleporting quantum states to one another. Human performers interact in real time with agents whose melodic and rhythmic behaviours are encoded as quantum states using Single Qubit Probability Amplitude Modulation (SQPAM)  and structured through Quantum Phase Estimation (QPE). Up to three agents are combined within a single quantum circuit, with directed communication via quantum teleportation. We are interested in supporting ambiguous, transformative interactions reminiscent of free Jazz improvisation. Therefore, rather than treating noise and decoherence as limitations, the system embraces NISQ-era constraints as creative affordances, framing agent communication as \textit{quantum whispers}, that is, deliberate, musically expressive imperfections in state transfer. We provide demonstrations and analyses based on melodic correlation, pitch-set distance, and state fidelity, where a continuum between imitation and divergence can be observed. We developed a tunable interpretation method to assess how agents reinterpret teleported states. This work positions teleportation as a promising interaction mechanism for agent-based quantum computer music and outlines future directions toward distributed ensembles connected via the Quantum Internet.

\medskip

This is a preprint of a paper accepted on 11 July 2026 for publication by Taylor \& Francis Group in the \textit{International Journal of Parallel, Emergent \& Distributed Systems}. Final version available at http://dx.doi.org/10.1080/17445760.2026.2703600}

\section{Introduction}
\label{sec:introduction}

Quantum computing has recently emerged as a novel computational paradigm for musicology and music technology, offering different methods for the representation and processing of music compared to classical digital systems \cite{Miranda2022a}. Quantum computing is unlocking new affordances for computer music systems by harnessing principles from quantum mechanics, such as superposition, entanglement and interference. Here, we are interested in developing systems to support interactive music-making, focusing on scenarios where human musicians and quantum musical agents make music together.

 \medskip
 
 The research reported in this paper explores the design of musical quantum agents, hereafter referred to simply as \textit{agents}. These are computational entities that process quantum information to generate musical material in response to incoming music. Conceptually, they listen to music played by one or more musicians, encode musical features into quantum states, and transform those states through quantum algorithms to produce musical responses. 
 
  \medskip
  
 Inter-agent communication is achieved using quantum teleportation  \cite{Bennett1993}. Teleportation-based interaction allows agents to influence the internal quantum states of one another during performance. That is, they communicate quantum states to each other. 

 \medskip
 
From a musical standpoint, we are keen to leverage NISQ (Noisy Intermediate-Scale Quantum) quantum computers for creativity. NISQ quantum computers are characterised by limited qubit counts, short coherence times, and significant error rates. Rather than treating these limitations as obstacles, we embrace them as creative affordances. In this context, we introduce the notion of \textit{quantum whisper} as a metaphor for agent communication in which divergence and ambiguity are tolerated. This approach draws inspiration from group improvisation practices, especially in free Jazz improvisation, where performers continuously interpret, misinterpret, and transform musical information they hear on the fly rather than reproducing exact copies of it.

 \medskip
 
However, whereas the present work focuses on circuits executed locally on currently available NISQ quantum devices, or simulators thereof, our long-term objective is more ambitious. Our research is aimed at future fault-tolerant quantum processors and large-scale quantum networks. We envision scenarios where agents are geographically distributed and connected through the emerging Quantum Internet \cite{Kimble2008, Hamdoun2020, Rohde2021}. Each agent would host its own quantum circuit embodying particular musical skills or behaviours. During a performance, quantum states representing such skills and behaviors would be teleported between distant agents, enabling collective composition, mutual adaptation, and learning across space in fundamentally distinctive ways from classical networked musical performance systems \cite{Gabrielli2016}. Networked musical performance today is constrained by classical communication bottlenecks, including latency, bandwidth limitations, and synchronisation issues. By exploring teleportation, this research contributes toward a longer-term vision of quantum-enhanced networked music systems that may overcome or reframe such constraints.

\medskip
 
As a frame of reference, several distributed agent systems for musical composition and improvisation have been developed before our quantum teleportation approach. Most notably, we cite the systems Voyager \cite{Lewis2000}, Cypher \cite{Rowe1992}, and GenJam \cite{Biles2007}. In the quantum computing domain, QuiKo and Lineage \cite{Oshiro2022, Oshiro2023} informed the foundations for the method of encoding music information into quantum circuits developed below. As an earlier attempt at deploying the teleportation protocol, please refer to \cite{Kirke2020}, also from ICCMR.

\medskip

The remainder of this paper is organised as follows. In Section \ref{sec:workflow}, we outline the overall system pipeline, while in Section \ref{sec:single_agent_circuit} we introduce the design of a single quantum musical agent, detailing the quantum circuit architecture, the encoding of musical features into quantum states, and the decoding of measurement results into musical output. Next, Section \ref{sec:multi-agents-demo} extends this architecture to multi-agent systems, presenting two- and three-agent configurations and describing how quantum teleportation mediates directed influence between agents. This section also reports experimental results obtained on both simulators and real quantum hardware, supported by musical and quantitative analyses. Then, in Section \ref{sec:monte_carlo}, we present an interpretation method to generalise the teleportation correction stage, enabling controlled variation in how agents reinterpret teleported states. Finally, Section \ref{sec:conclusion} concludes with a discussion of the musical and computational implications of teleportation-based interaction. We propose future directions for this research, toward larger groups and geographically distributed agents.

\section{System Pipeline}
\label{sec:workflow}

\begin{figure}[htbp]
\begin{center}\vspace{0.3cm}
\includegraphics[width=0.4\linewidth]{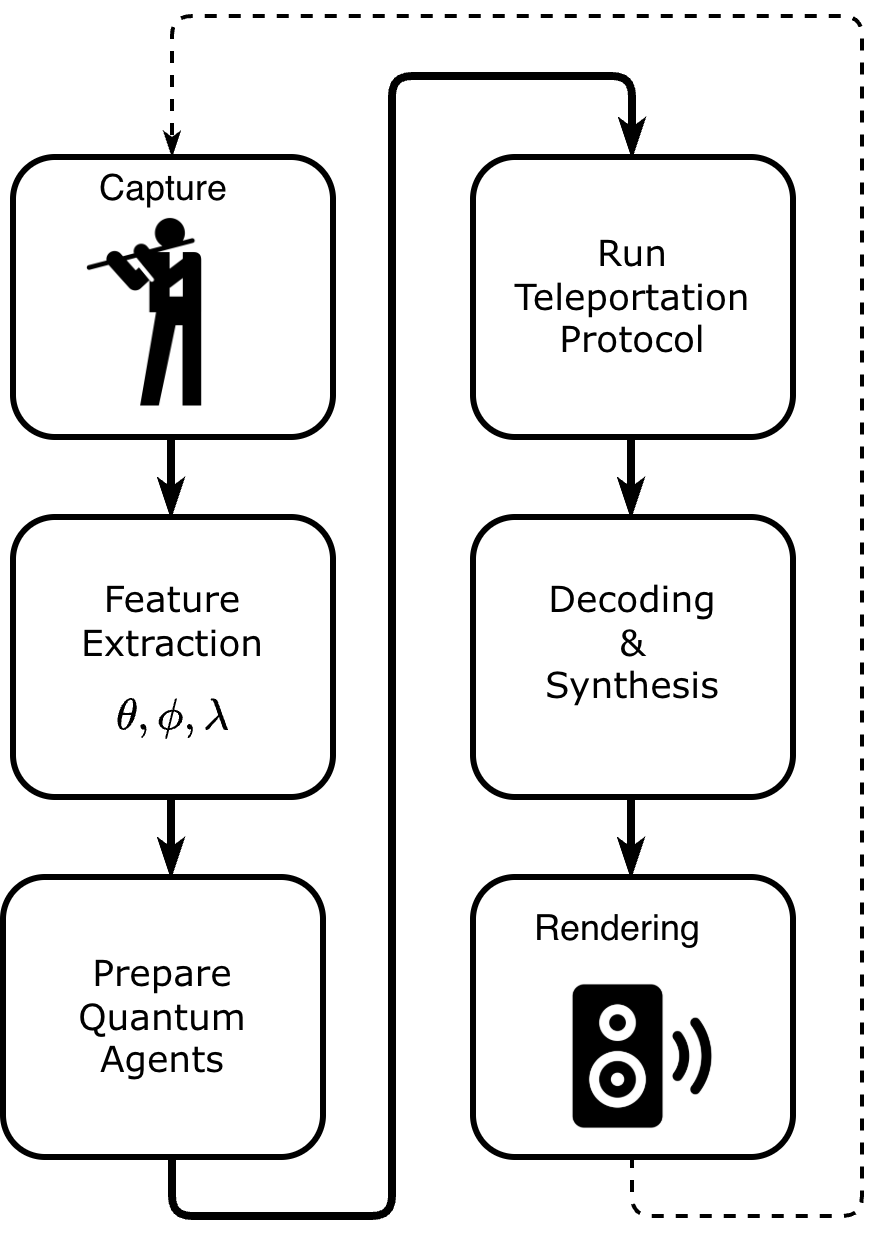}
\caption{Quantum Agent System Pipeline.}
\label{fig:Workflow}
\end{center}
\end{figure}

The overall workflow of the quantum agent system is illustrated in Fig. \ref{fig:Workflow}. An iteration within the system begins with the capture of a live audio stream and the extraction of its audio features. These features serve as parameters for a quantum circuit embodying the agents. Then, the system prepares the agents (i.e., builds the quantum circuit), and the quantum teleportation protocol is executed, transferring the agent's state to another agent in the system. Typically, the circuit is run over multiple shots (repeated executions of the same quantum circuit) to gather statistically significant data and mitigate errors caused by hardware noise. However, noise can be tolerated to our advantage, as discussed briefly in the Introduction. Finally, the results from running the circuit are decoded into melodic and rhythmic information, which are subsequently rendered into musical output for playback. After, or even during, playback, the system is ready for another iteration.

\medskip

Technically, the system employs the Single Qubit Probability Amplitude Modulation (SQPAM) representation scheme \cite{Itaborai2022}, combined with the Quantum Phase Estimation (QPE) algorithm \cite{Ueda2024}.  Melodic and rhythmic information extracted from input music is encoded into the probability amplitudes of individual qubits with SQPAM. Then, QPE is used to entangle this information with a representation of time, allowing the agents to generate structured musical sequences. Upon measurement of the respective quantum circuit, the results are decoded and rendered as music, which can be played using MIDI software instruments \cite{MIDI_Association2025} (Fig. \ref{fig:Workflow}).

\section{Single Quantum Agent}
\label{sec:single_agent_circuit}

\begin{figure}[H]
\begin{center}\vspace{0.3cm}
    \begin{tikzpicture}
        \node[scale=0.6] 
        {
            \begin{quantikz}
                	\lstick{$q_0$} & \ket{0} &  \gate{\textbf{H}} & \gate{\textbf{X}}  & \ctrl{3}   & \hphantomgate{} & \hphantomgate{}	& \ctrl{4}  & \hphantomgate{} & \hphantomgate{} &  \qw & \ldots \\
                	\lstick{$q_1$} & \ket{0} &  \gate{\textbf{H}} & \gate{\textbf{X}}  & \hphantomgate{} & \ctrl{2}  & \hphantomgate{} & \hphantomgate{} & \ctrl{3}  & \hphantomgate{}	 & \qw & \ldots \\
                	\lstick{$q_2$} & \ket{0} &  \gate{\textbf{H}} & \gate{\textbf{X}}  & \hphantomgate{} & \hphantomgate{} & \ctrl{1}  & \hphantomgate{} &  \hphantomgate{} & \ctrl{2}  & \qw & \ldots \\[1cm]
                	\lstick{$q_3$} & \ket{0} &  \hphantomgate{} &  \hphantomgate{}  & \gate[style={fill=yellow!30}]{\textbf{U}(\theta, \phi, \lambda)} & \gate[style={fill=yellow!30}]{\textbf{U}(\theta, \phi, \lambda)} & \gate[style={fill=yellow!30}]{\textbf{U}(\theta, \phi, \lambda)} & \hphantomgate{} &  \hphantomgate{} & \hphantomgate{} & \qw & \ldots \\
	    	\lstick{$q_4$} & \ket{0} &  \hphantomgate{} & \hphantomgate{}   & \hphantomgate{} &  \hphantomgate{} & \hphantomgate{}  & \gate[style={fill=yellow!30}]{\textbf{U}(\theta, \phi, \lambda)} & \gate[style={fill=yellow!30}]{\textbf{U}(\theta, \phi, \lambda)} & \gate[style={fill=yellow!30}]{\textbf{U}(\theta, \phi, \lambda)} & \qw & \ldots \\[2cm]		
		\lstick{$q_0$} \ldots & & \gate{\textbf{X}} & \ctrl{3}   & \hphantomgate{} & \hphantomgate{}	& \ctrl{4}  & \hphantomgate{} & \hphantomgate{} & \gate[3, style={fill=red!30}][2cm]{\textbf{QFT}^{-1}} & \meter{} &  \\
                	\lstick{$q_1$} \ldots & & \gate{\textbf{X}} & \hphantomgate{} & \ctrl{2}  & \hphantomgate{} & \hphantomgate{} & \ctrl{3}  & \hphantomgate{} & \hphantomgate{} & \meter{}  &  \\
                	\lstick{$q_2$} \ldots & & \gate{\textbf{X}}  & \hphantomgate{} & \hphantomgate{} & \ctrl{1}  & \hphantomgate{} &  \hphantomgate{} & \ctrl{2}  & \hphantomgate{} & \meter{} & \\[1cm]
                	\lstick{$q_3$}  \ldots& & \hphantomgate{} & \gate[style={fill=yellow!30}]{\textbf{U}(\theta, \phi, \lambda)} & \gate[style={fill=yellow!30}]{\textbf{U}(\theta, \phi, \lambda)} & \gate[style={fill=yellow!30}]{\textbf{U}(\theta, \phi, \lambda)} & \hphantomgate{} &  \hphantomgate{} & \hphantomgate{}  & \hphantomgate{} & \meter{} & \\
	    	\lstick{$q_4$}  \ldots & & \hphantomgate{} & \hphantomgate{} &  \hphantomgate{} & \hphantomgate{}  & \gate[style={fill=yellow!30}]{\textbf{U}(\theta, \phi, \lambda)} & \gate[style={fill=yellow!30}]{\textbf{U}(\theta, \phi, \lambda)} & \gate[style={fill=yellow!30}]{\textbf{U}(\theta, \phi, \lambda)} & \hphantomgate{} & \meter{}  &  \

            \end{quantikz} 
        };
    \end{tikzpicture}
\end{center}
\caption{Quantum Circuit of a single quantum agent.}
\label{fig:Agent_Structure}
\end{figure}
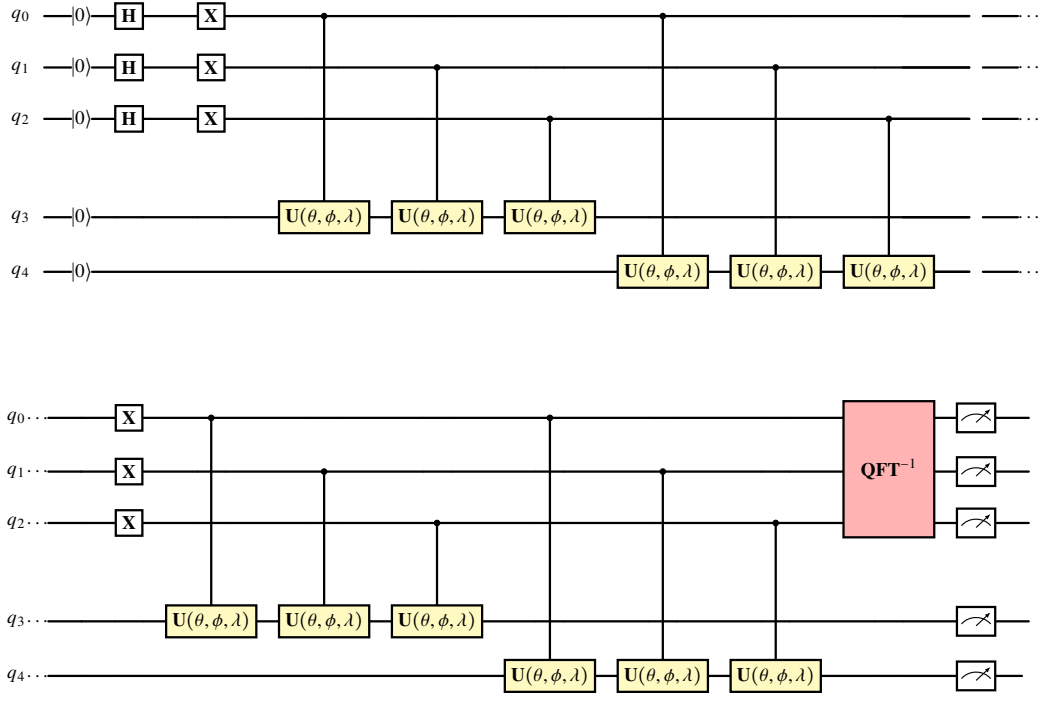

Before discussing the full multi-agent system, let us first introduce the design of a single quantum agent. Consider the case where an agent is designed to generate 8-note sequences. The agent is implemented using the quantum circuit shown in Fig. \ref{fig:Agent_Structure}.

The circuit implements the Single Qubit Probability Amplitude Modulation (SQPAM) scheme \cite{Itaborai2022}, and the Phase Kickback Sequencing Encoding (PKBSE) \cite{Oshiro2022} method. SQPAM is a representation scheme originally designed for the representation of digital audio signals on quantum circuits. It requires:

\begin{itemize}
\item{An amplitude qubit: A single qubit that stores the normalised amplitude information via its probability amplitudes (i.e., the probabilities of measuring $\ket{0}$ or $\ket{1}$).}
\item{A time register: A set of $n$ qubits that represent the time indices - or positions - in the signal sequence, where the total signal length is $N = 2^n$.}
\end{itemize}

\medskip

One set of qubits ($q_0$, $q_1$ and $q_2$) represents the time register. Each of the binary codes produced by the three-qubit register corresponds to eighth-note subdivisions in a $\frac{4}{4}$ time signature. The remaining two qubits ($q3$ and $q4$), referred to as signal qubits, represent the melodic (pitch) and rhythmic (duration) information at each subdivision.

\medskip

Through trigonometric mappings of rotation gate parameters, SQPAM modulates the probabilities of a single qubit to represent the amplitude of the signal at each time step. This allows the amplitude information of the entire signal to be represented as a quantum superposition. Unlike other multi-qubit encoding schemes (e.g., basis encoding, which requires $\log_2(N)$ qubits for $N$ amplitude values), SQPAM uses just one qubit for all amplitudes, making it resource-efficient. 

\medskip

The PKBSE method is utilised in this design to encode musical features. As outlined in the system pipeline, each quantum agent will extract features from the acoustic signal generated by external live performers. These features will then be used as parameters for controlled Unitary gates  \textbf{U}($\theta$, $\phi$, $\lambda$) that entangle the time register qubits with the signal qubits. In our implementation, we use Qiskit's \code{U3Gate} \cite{Qiskit2025}, referred to in this paper simply as the \textbf{U} gate. The \textbf{U} gate requires three Euler angles $\theta$, $\phi$, and $\lambda$ to rotate the state vector of a qubit. Mathematically, the \textbf{U} gate is defined as follows (Eq. \ref{eq:unitary_equation}):

\begin{equation}
\normalsize
    U(\theta, \phi, \lambda) = \begin{pmatrix}
    	cos \frac{\theta}{2}  & -e^i \lambda sin \frac{\theta}{2} \\ \\
	e^i \phi sin \frac{\theta}{2}  & e^i (\phi + \lambda) cos \frac{\theta}{2}
	\end{pmatrix}
    \label{eq:unitary_equation}
\end{equation}

\noindent
Commonly, multi-controlled \textbf{U} gates are used to entangle the two registers to keep the rotations associated with the time indices they occur on. This is seen in a number of quantum audio signal representations \cite{Itaborai2022}. This is not efficient because multi-controlled \textbf{U} gates decompose into a set of single-qubit and single-control qubit gates. If we introduce more time qubits into the circuit,  the circuit depth would increase exponentially. This increases the chances of decoherence of the overall quantum state. PKBSE is able to associate the specific rotations on the signal qubits with the correct time index (subdivision), implementing only single-controlled \textbf{U} gates. The PKBSE method is detailed in \cite{Oshiro2022}.

\medskip
At its core, PKBSE utilises the Quantum Phase Estimation (QPE) algorithm \cite{Ueda2024} to exploit the phase kickback induced by the controlled \textbf{U} gates. In essence, the phases from the controlled \textbf{U} gates are applied to the signal qubits and are \enquote{kicked back} to the time register via this phenomenon. This introduces temporal variability, potentially shifting the subdivision where pitch or duration content manifests. For instance, a phase $\phi = \frac{3}{4}\pi$ applied to one of the signal qubits (e.g., melodic qubit) when $q_0$ is an excited state (i.e. $\ket{1}$), bestows that phase to the signal qubit, but also kicks back and applies the phase to $q_0$. Then, the inverse Quantum Fourier Transform (QFT) \cite{Bernhardt2019} converts these phase-basis rotations to the computational basis for measurement. In our system, $\phi$ and $\lambda$, shown in Eq. \eqref{eq:unitary_equation}, represent the phase angles being kicked back to the time register qubits. 

\subsection{Feature Extraction}
\label{sec:feature_extraction}

As discussed in the previous subsection, the Euler angles for our controlled \textbf{U} gates are defined from musical features extracted from an input audio stream being generated by a live performer. Several strategies exist for segmenting the audio stream and extracting features, such as channel separation, filtering, or onset-based windowing \cite{Giannakopoulos2014}. Our system extracts these features using the Librosa \cite{McFee2015} Python library. In the present implementation, we extract the following features:

\begin{itemize}
	\item{The contour energy of the melodic phrase for $\theta_m$}
	\item{The average tempo value of the phrase $\theta_r$}
	\item{The spectral centroid of the audio signal, for $\phi_m$ and $\phi_r$}
	\item{The onset intensity, for $\lambda_m$ and $\lambda_r$}
\end{itemize}

Subscript $m$ indicates that the feature parameter is associated with the melodic qubit, while subscript $r$ indicates that it is a feature parameter associated with the rhythmic qubit.

\medskip

The contour energy of the audio signal is calculated utilising the Probabilistic YIN (pYIN) algorithm \cite{Mauch2014} in Librosa to pitch-track the input audio signal into the system. The audio stream is segmented into 3 equal blocks, with each block being associated with each qubit in the time register. Each block is entangled with the melodic and rhythmic qubits. The \code{pYIN} function from Librosa is applied to calculate an array of fundamental frequencies $f_0$ at different time steps throughout each block, and then converted to MIDI values. From here, the sum of all the intervals between neighbouring pitches in the block is then divided by the block size. This approach is inspired by a method for calculating melodic correlation introduced in \cite{Kim2024}, which will be used later in sections to compare generated melodies between agents. Contour energy can be expressed in Eq. \ref{eq:energy-contour}.

\begin{equation}
	\normalsize
	CE = \frac{\sum_{i=1}^{N} {\lvert P_{i-1}- P_{i}}\rvert}{B_s}
	\label{eq:energy-contour}
\end{equation}

$CE$ represents contour energy, while $P$ represent the pitch array of MIDI values converted from the fundamental frequencies from pYin. $N$ represents the number of pitches (notes) present within $P$, and $B_s$ represents the Block size of the audio segment.We start on the second element of the pitch array and subtract from the previous pitch MIDI value to obtain the interval between the neighbouring pitches. If the interval values are relatively close to one another and consistently moving in the same direction (positive up (higher) or negative, down (lower)), $CE$ is small. If the intervals between notes increase with varying direction, then $CE$ increases. This feature will be used as a parameter for the $\theta_m$ on the melodic qubit. This is a useful metric for interacting with quantum systems using external stimuli in that the quantum state becomes sensitive to the melodic shape of the live performer. 

\medskip

Parameter $\theta_r$ represents the rhythmic feature to be encoded onto the rhythmic signal qubit of the quantum agent circuit. Here, we choose to extract and encode the average tempo over time of the input audio signal. This is calculated in Librosa using the \code{beat.tempo()} function with the onset envelope of the audio signal as its input.  In addition, for both signal qubits, the phases, represented as parameters $\phi_m$, $\phi_m$, $\lambda_m$ and $\lambda_r$, spectral centroid \cite{Schubert2004} and onset intensity features from the input audio signal are used. These functions come directly from Librosa in \code{librosa.onset.onset\_strength()} and \code{librosa.feature.spectral\_centroid()}. Spectral Centroid measures the centre of mass of a signal, and onset intensity measures the intensity of the transients within the signal. 

\medskip

Tables \ref{tab:trace_mel} and \ref{tab:trace_rhy} show the results of the feature extraction process of running an audio signal of a flute performance through the different extraction functions. These features are encoded onto the controlled \textbf{U} gates as outlined in the previous subsection. Each feature is scaled to be within values of -10 to +10 to prevent the features from becoming too large, while being able to move the qubit state throughout at least one rotation of the Bloch sphere, $0 < \theta < 2\pi$. In the following subsection, results obtained from running the single-agent circuit, with the features and parameters presented in this subsection, on the Qiskit Aer simulator and on real quantum hardware will be presented and discussed.

\begin{table}[htbp]
    \centering
    \caption{Melodic signal qubit feature extraction and conversion to \textbf{U} gate parameter.}
    \label{tab:trace_mel}
    \begin{tabular}{|c | c | c | c | c | c | c | c |}
        \toprule
        \textbf{Qubit} & \textbf{Function} &  \boldmath $\theta_{m0}$ &  \boldmath $\phi_{m0}$ &  \boldmath $\lambda_{m0}$ &   \boldmath $\theta_{m1}$ &  \boldmath $\phi_{m1}$ &  \boldmath $\lambda_{m1}$ \\
        \midrule
        $q_0$ & $contour\_energy( data )$ & -0.307& & &-0.002& &\\
        $q_0$ & $ spectral\_centroid( data )$ & & 7.888 & & &4.155 &\\
        $q_0$ & $ onset\_intensity( data )$ & & & 1.752 & & & 1.353\\
        
        $q_1$ & $ contour\_energy( data )$ & 0.132& & &0.127& &\\
        $q_1$ & $ spectral\_centroid( data )$ & & 4.436 & & &4.459 &\\
        $q_1$ & $ onset\_intensity( data )$ & & & 1.231& & & 1.420\\
        
        $q_2$ & $ contour\_energy( data )$ & -0.033& & &-3.656& &\\
        $q_2$ & $ spectral\_centroid( data )$ & & 4.913 & & &5.568 &\\
        $q_2$ & $ onset\_intensity( data )$ & & & 1.424 & & & 1.672\\
         \bottomrule
    \end{tabular}
\end{table}

\begin{table}[htbp]
    \centering
    \caption{Rhythmic signal qubit feature extraction and conversion to \textbf{U} gate parameter.}
    \label{tab:trace_rhy}
    \begin{tabular}{|c | c | c | c | c | c | c | c |}
        \toprule
        \textbf{Qubit} & \textbf{Function} &  \boldmath $\theta_{r0}$ &  \boldmath $\phi_{r0}$ &  \boldmath $\lambda_{r0}$ &   \boldmath $\theta_{r1}$ &  \boldmath $\phi_{r1}$ &  \boldmath $\lambda_{r1}$ \\
        \midrule
        $q_0$ & $ tempo( data )$ & 1.58 & & &0.360& &\\
        $q_0$ & $ spectral\_centroid( data )$ & & 7.888 & & &4.155 &\\
        $q_0$ & $ onset\_intensity( data )$ & & & 1.752 & & & 1.353\\
        
        $q_1$ & $ tempo( data )$ & 1.153& & &0.360& &\\
        $q_1$ & $ spectral\_centroid( data )$ & & 4.436 & & &4.459 &\\
        $q_1$ & $ onset\_intensity( data )$ & & & 1.231& & & 1.420\\
        
        $q_2$ & $ tempo( data )$ & 0.123& & &0.123& &\\
        $q_2$ & $ spectral\_centroid( data )$ & & 4.913 & & &5.568 &\\
        $q_2$ & $ onset\_intensity( data )$ & & & 1.424 & & & 1.672\\
         \bottomrule
    \end{tabular}
\end{table}

\subsection{ Results and Decoding}

Figures \ref{fig:single_agent_results_sim} and \ref{fig:single_agent_Results_ibmfez} below show the probability distribution results from running the single-agent circuit on both Qiskit's Aer simulator and on quantum IBM hardware, ibm\_fez.

\begin{figure}[htbp]
    \centering
    \subfigure[]{
       \includegraphics[scale=0.575]{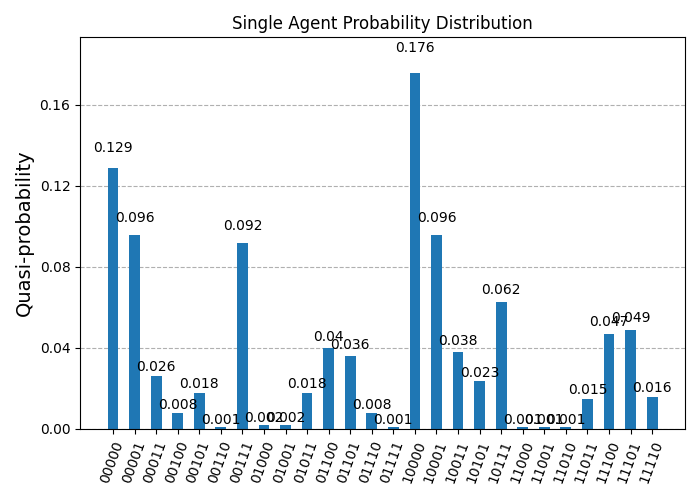}
       \label{fig:single_agent_results_sim}
    }
    \subfigure[]{
      \includegraphics[scale=0.575]{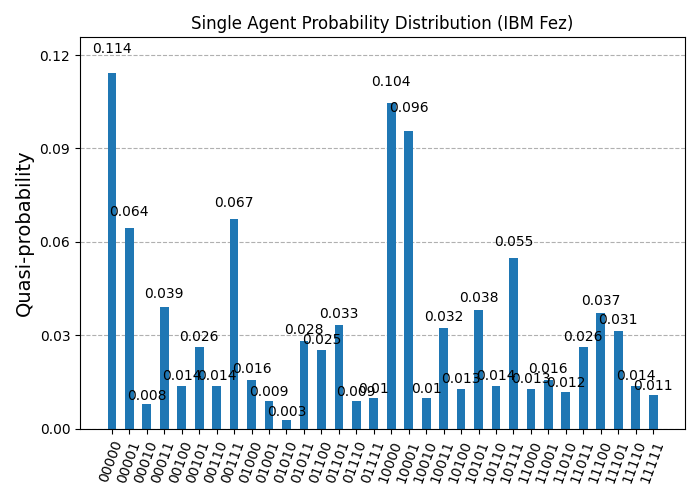}
       \label{fig:single_agent_Results_ibmfez}
    }
    \caption{Probability distribution for a single quantum agent produced on (a) Qiskit's Aer simulator and (b) ibm\_fez.}
    \label{fig:single_agent_prob_dist}
\end{figure}

After obtaining the distributions presented in Fig. \ref{fig:single_agent_prob_dist} from measuring the system, the SQPAM-encoded pitch and duration information can now be decoded and rendered into complete melodic phrases. Eq. \ref{eq:decoding}, as presented in \cite{Itaborai2022}, gives us a way to decode out the melodic and rhythmic information with respect to time. In Eq. \ref{eq:decoding}, $p_i^{|1\rangle}$ denotes the probability of measuring the qubit in the $|1\rangle$ state for subdivision $i$, and $p_i^{|0\rangle}$ is the probability of measuring the qubit in the $|0\rangle$ state. 

\begin{equation}
	a_i = \frac{2 p_i ^ {\ket{1}}} {p_i ^ {\ket{0}} + p_i ^ {\ket{1}}} -1
	\label{eq:decoding}
\end{equation}

Since ${p_i ^ {\ket{0}} + p_i ^ {\ket{1}}} =1$, Eq. \ref{eq:decoding} simplifies to $a_i = 2 p_i^{|1\rangle} - 1$, where $a_i$ represents the amplitude associated with subdivision $i$. For each binary string corresponding to a measured state of qubits $q_0$, $q_1$, and $q_2$, which defines the subdivision, the qubits for pitch and duration ($q_3$, $q_4$) yield a probability distribution over $\ket{0}$  and $\ket{1}$. The resulting decoded signals can then be mapped to musical patterns for playback via a MIDI instrument or converted to a notated musical score. The demonstration results are illustrated in the results below in Fig. \ref{fig:singleagent_dec_res_sim} and  \ref{fig:singleagent_dec_res_real}. 

\medskip

In rendering the complete melodic phrase from the decoded melodic and rhythmic signals, each float value in the sequences is quantised to pitches within a chosen scale. The scores generated below are in the key of F major. However, the quantum agent could potentially move between different scales and pitch sets as done in \cite{Biles2007}. In the rhythmic signal, we picked a reference tempo (e.g., 120 bpm) and then multiplied it by the float values in the rhythmic signal. This created a more flexible interpretation of the rhythm of the resulting musical phrase rather than just a quantised set of note durations (e.g., whole, half, quarter, eighth, or sixteenth notes). 

\begin{figure}[htbp]
    \centering
    \subfigure[Melodic and rhythmic qubit signals of \texttt{agent 1}.]{
       \includegraphics[scale=0.6]{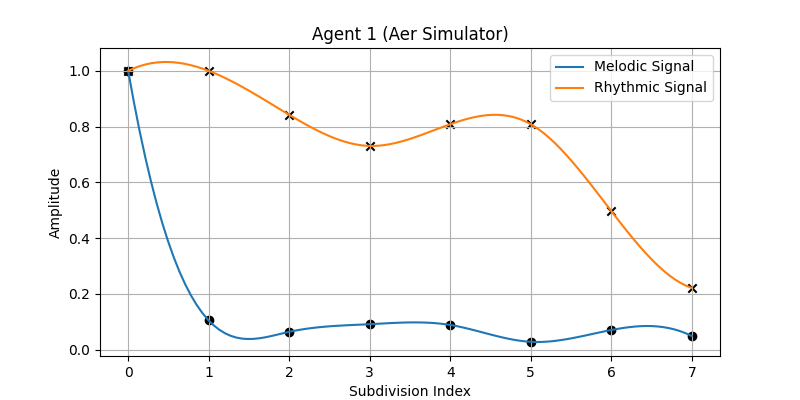}
       \label{sub:Single_agent_sim_plot}
    }
    \subfigure[Rendered musical notation of \texttt{agent 1}.]{
      \includegraphics[scale=0.75]{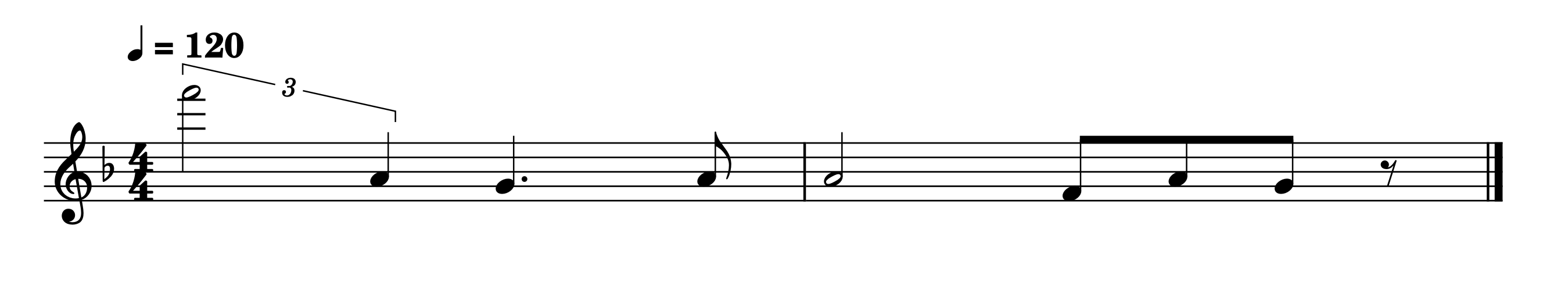}
       \label{sub:Single_agent_sim_music}
   }
    \caption{Single agent (a) melodic and rhythmic qubit signals and (b) decoded music notation produced with Aer simulator.}
    \label{fig:singleagent_dec_res_sim}
\end{figure}

\begin{figure}[htbp]
    \centering
    \subfigure[Melodic and rhythmic qubit signals of \texttt{agent 1}.]{
       \includegraphics[scale=0.6]{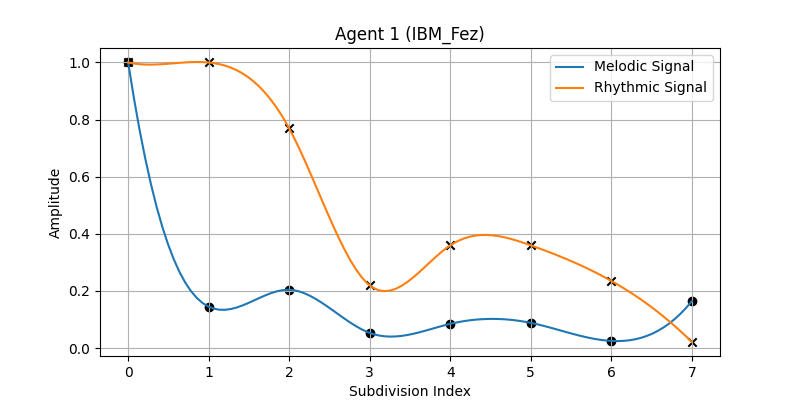}
       \label{sub:Single_agent_Fez_plot}
    }
    \subfigure[Rendered musical notation of \texttt{agent 1}.]{
       \includegraphics[scale=0.75]{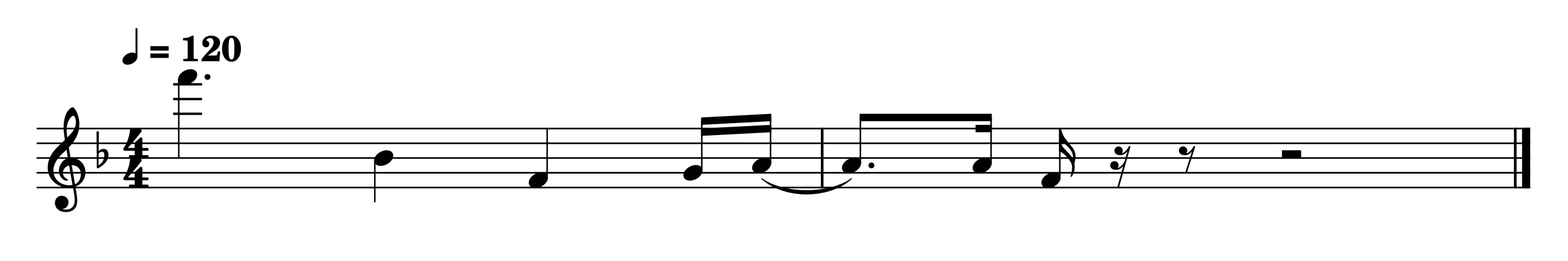}
      \label{sub:Single_agent_Fez_music}
    }
    \caption{Single agent (a) melodic and rhythmic qubit signals and (b) decoded music notation produced with the IBM quantum hardware, ibm\_fez.}
    \label{fig:singleagent_dec_res_real}
\end{figure}

\section{Multiple Quantum Agents with Teleportation}
\label{sec:multi-agents-demo}

In this section, we present practical demonstrations designed to illustrate how the agents introduced above behave and interact within a multi-agent system. Specifically, this section looks at musical results using two and three agents. These qualitative analyses are used to showcase instances where quantum teleportation can mediate musical influence between multiple agents, and how increasing the number of agents and teleportation pathways affects musical structure, variability, and interpretability. We can use these examples as a way to begin developing testbeds in which musicians and audiences can study the behaviour and performance of these systems by listening and interacting with them. It takes advantage of a more physical and organic mode of understanding than a statistical analysis.

\medskip
For both demonstrations, the same core pipeline described in Section \ref{sec:workflow} is employed. These demonstrations should, however, be understood not as benchmarks of computational efficiency, but as explorations of musically meaningful behaviour under NISQ constraints. The metrics calculated here are qualitative, in that they act as observations that we can use as a starting point in gaining an intuition of how these agents interact with one another and human performers. All experiments were run both on Qiskit’s Aer simulator and on IBM quantum processors (\code{ibm\_kingston} and \code{ibm\_fez}), allowing us to contrast idealised behaviour with hardware-induced noise. However, before diving into the demonstrations of the multi-agent system, a discussion on the inner workings of the quantum teleportation protocol in the context of our architecture is presented.

\begin{figure}[htbp]
\begin{center}\vspace{0.3cm}
    \begin{tikzpicture}
        \node[scale=0.75] 
        {
            \begin{quantikz}
            
                	\lstick{$q_0$} & \gate[5, style={fill=yellow!30}]{\textbf{  Agent 1  }} \slice{One Agent} & \hphantomgate{} & \hphantomgate{} \slice{Two Agents} &  \hphantomgate{} & \hphantomgate{}   \slice{Three Agents} & \hphantomgate{}   \\
                	\lstick{$q_1$} &  \hphantomgate{} & \hphantomgate{} & \hphantomgate{} &  \hphantomgate{} & \hphantomgate{}  & \hphantomgate{}  \\
	 	\lstick{$q_2$} & \hphantomgate{} & \hphantomgate{} 	& \hphantomgate{} &  \hphantomgate{} & \hphantomgate{} & \hphantomgate{} \\[0.5cm]
		\lstick {Melody $q_3$} & \hphantomgate{}  & \gate[7, style={fill=red!30}]{Teleportation} &  \hphantomgate{} & \hphantomgate{}  & \hphantomgate{} & \hphantomgate{} \\
                	\lstick {Rhythm $q_4 $} & \hphantomgate{} & \hphantomgate{} 	& \hphantomgate{} &  \hphantomgate{} & \hphantomgate{}  &  \hphantomgate{} \\[0.8cm]
	
	\lstick {$aux_1$ $q_5 $} & \hphantomgate{} & \hphantomgate{} 	& \hphantomgate{} &  \hphantomgate{} & \hphantomgate{}  &  \hphantomgate{} \\[0.8cm]
	
		\lstick{$q_6$} & \hphantomgate{} &  \hphantomgate{} & \gate[5, style={fill=blue!30}]{\textbf{  Agent 2  }} & \hphantomgate{}  & \hphantomgate{} & \hphantomgate{} 	\\
	 	\lstick{$q_7$} & \hphantomgate{} & \hphantomgate{} 	& \hphantomgate{} &  \hphantomgate{} & \hphantomgate{}  & \hphantomgate{}   \\
		\lstick{$q_8$} & \hphantomgate{} & \hphantomgate{} 	& \hphantomgate{} &  \hphantomgate{} & \hphantomgate{}  & \hphantomgate{}   \\[0.5cm]
		\lstick{Melody $q_9$} & \hphantomgate{} & \hphantomgate{} &  \hphantomgate{}  & \gate[7, style={fill=red!30}]{Teleportation} &  \hphantomgate{} & \hphantomgate{}  \\
                	\lstick{Rhythm $q_{10}$} & \hphantomgate{} & \hphantomgate{} 	& \hphantomgate{} & \hphantomgate{} & \hphantomgate{} & \hphantomgate{}   \\[0.8cm]
	
	\lstick {$aux_2$ $q_{11} $} & \hphantomgate{} & \hphantomgate{} 	& \hphantomgate{} &  \hphantomgate{} & \hphantomgate{}  &  \hphantomgate{} \\[0.8cm]

		\lstick{$q_{12}$}   & \hphantomgate{} & \hphantomgate{} & \hphantomgate{} 	&  \hphantomgate{} & \gate[5, style={fill=green!30}]{\textbf{  Agent 3  }} & \hphantomgate{}  \\
	 	\lstick{$q_{13}$} & \hphantomgate{} & \hphantomgate{} & \hphantomgate{} 	&  \hphantomgate{} &  \hphantomgate{} & \hphantomgate{}  \\
		\lstick{$q_{14}$} & \hphantomgate{} & \hphantomgate{} & \hphantomgate{} 	&  \hphantomgate{} &  \hphantomgate{} & \hphantomgate{} \\[0.5cm]
		\lstick{Melody $q_{15}$} & \hphantomgate{} & \hphantomgate{} & \hphantomgate{} &  \hphantomgate{} &  \hphantomgate{} & \hphantomgate{} \\
                	\lstick{Rhythm $q_{16}$} & \hphantomgate{} & \hphantomgate{} & \hphantomgate{} &  \hphantomgate{} &  \hphantomgate{}  & \hphantomgate{}  \	
			
            \end{quantikz} 
        };
    \end{tikzpicture}
\end{center}
\caption{Multiagent scheme with teleportation.}
\label{fig:Multiagent_Circuit}
\end{figure}
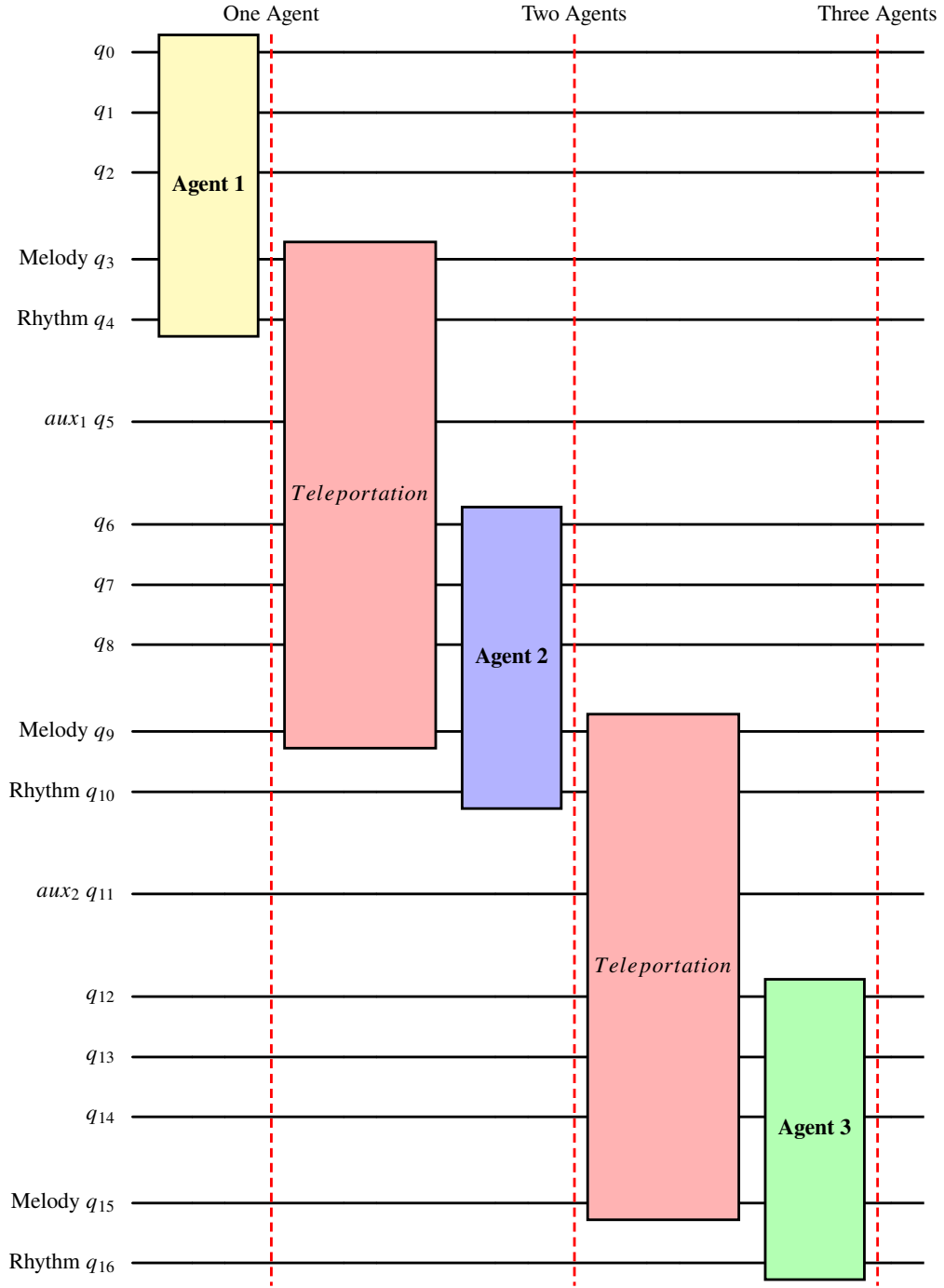

\subsection{The Teleportation Protocol}
\label{sec:teleport}

\begin{figure}[htbp]
\begin{center}\vspace{0.3cm}
    \begin{tikzpicture}
        \node[scale=0.9] 
        {
            \begin{quantikz}
                	\lstick{$q_0$} & \ket{\Psi}_1 & \gate[style={fill=yellow!30}]{\textbf{U}(\theta, \phi, \lambda)} & \hphantomgate{} & \ctrl{1} & \gate{\textbf{H}}  & \meter{}   \\
                	\lstick{$q_1$} & \ket{0} & \gate{\textbf{H}} & \ctrl{1} & \gate{\textbf{X}}  & \meter{}   \\
                	\lstick{$q_2$} & \ket{0} &  \hphantomgate{} & \gate{\textbf{X}} & \hphantomgate{} & \gate{\textbf{X}} \wire[u][1]{c} & \gate{\textbf{Z}} \wire[u][2 ]{c} &  \ket{\Psi}_2 & \meter{} \		
            \end{quantikz} 
        };
    \end{tikzpicture}
\end{center}
\caption{Quantum Teleportation Protocol Circuit}
\label{fig:Teleportation_Explanation}
\end{figure}
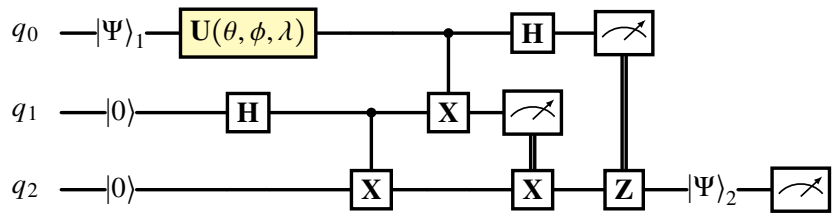

The teleportation protocol \cite{Bennett1993, Djordjevic2022, Hamdoun2020} enables the transfer of a quantum state from one agent to another, allowing the receiving agent to adopt the state of the sender as its own initial state before progressing through its own unique state evolution. Figure \ref{fig:Multiagent_Circuit} depicts a multi-agent quantum circuit, illustrating the cascading teleportation-based interactions among multiple agents within our system, and Fig. \ref{fig:Teleportation_Explanation} shows the teleportation scheme being implemented in our system. Here we consider a scenario where \texttt{agent 1} aims to communicate a quantum state $\ket{\Psi}_1$ to \texttt{agent 2}. The circuit presented in Fig. \ref{fig:Teleportation_Explanation} shows the teleportation of $\ket{\Psi}_1$ = \textbf{U}($\theta, \phi, \lambda$) on qubit $q_0$ to qubit 2 ($q_2$), resulting in $\ket{\Psi}_2$. \texttt{Agent 1} requires an auxiliary qubit $q_1$ (this is qubit $q_5$ and $q_{11}$ in Fig. \ref{fig:Multiagent_Circuit})  to entangle with and create a Bell state \cite{Djordjevic2022}. This is done using a Hadamard (\textbf{H}) gate followed by a controlled-X (a.k.a. controlled-NOT) (\textbf{CX}) gate. Then, $q_0$ and $q_1$ are entangled via another \textbf{CX} gate, after which an \textbf{H} gate is applied to $q_0$ to shift it into the Hadamard (phase) basis.

By measuring $q_0$ and $q_1$ we obtain a code determining in what basis $q_2$ should be measured to \enquote{receive} the state $\ket{\Psi}_2$:

\begin{itemize}
\item{A measurement outcome of $\ket{00}$ no correction gates on $q_2$ to recover $\ket{\Psi}_2$.}
\item{A measurement outcome of $\ket{01}$ requires a \textbf{Z} gate on $q_2$ to recover $\ket{\Psi}_2$.}
\item{A measurement outcome of $\ket{10}$ requires a \textbf{X} gate on $q_2$ to recover $\ket{\Psi}_2$.}
\item{A measurement outcome of $\ket{11}$ requires both gates \textbf{Z} and \textbf{X} on $q_2$ to recover $\ket{\Psi}_2$.}
\end{itemize}

\medskip

Teleportation forms a core building block in our multi-agent system. It enables direct influence between agents. The state of the receiver can evolve independently afterwards due to subsequent quantum gates,  decoherence, etc. This is significant in the sense that agents can influence each other at specific points within the time evolution of the system, but can then diverge or converge depending on each agent's individual design and parameters.

\subsection{Two Agents Example}
\label{sec:cas_two}

\medskip

The following example focuses on the first two agents in Fig. \ref{fig:Multiagent_Circuit}. Here, the incoming audio stream from the live musician was segmented into two equal parts. The first segment, representing the first half of the audio stream, was assigned to \texttt{agent 1}. The second segment, the second half of the audio stream, was assigned to \texttt{agent 2}.  Because these segments are two different audio chunks, they resulted in different values for the musical features encoded within each agent. That is, the quantum state of \texttt{agent 1} will be different from the quantum state of \texttt{agent 2}. 

\medskip

\begin{figure}[htbp]
    \centering
    \subfigure[Melodic and rhythmic qubit signals of \texttt{agent 1}.]{
         \includegraphics[scale=0.6]{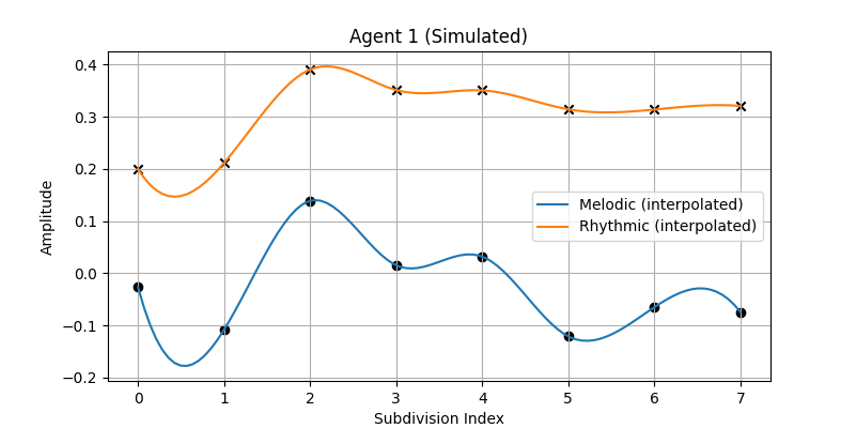}
        \label{fig:subim1}
    }
    \subfigure[Rendered musical notation of \texttt{agent 1}.]{
       \includegraphics[scale=0.75]{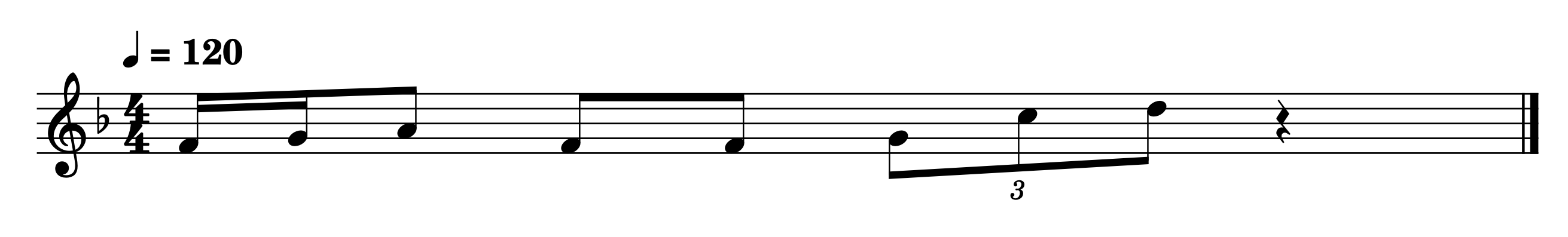}
        \label{fig:subim2}
    }
    \caption{Results for \texttt{agent 1} executed on Qiskit's Aer simulator.}
    \label{fig:cas_ag1_sim}
\end{figure}

\begin{figure}[htbp]
    \centering
    \subfigure[Melodic and rhythmic qubit signals of \texttt{agent 2}.]{
         \includegraphics[scale=0.6]{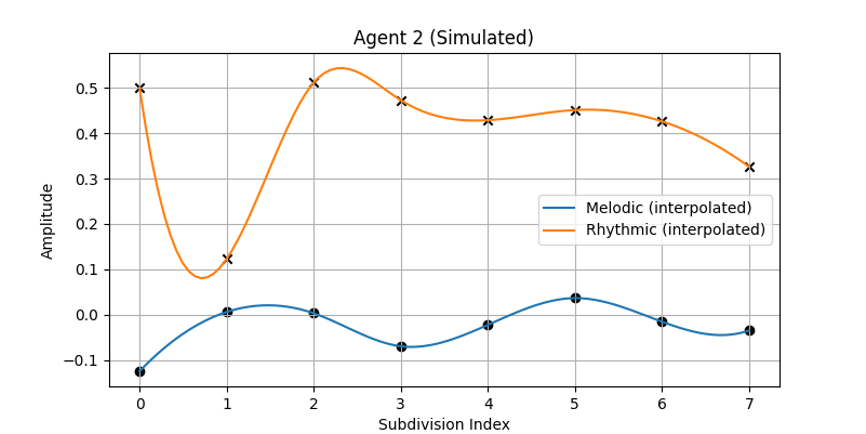}
        \label{fig:subim3}
    }
    \subfigure[Rendered musical notation of \texttt{agent 2}.]{
        \includegraphics[scale=0.75]{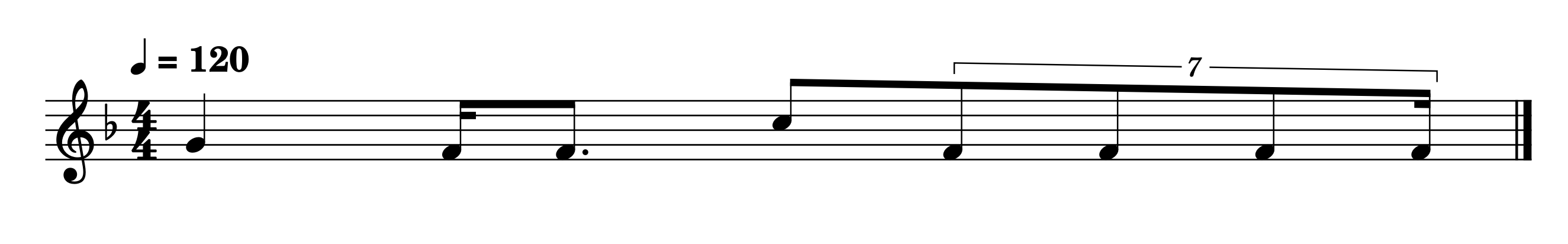}
        \label{fig:subim4}
    }
    \caption{Results for \texttt{agent 2} executed on Qiskit's Aer simulator.}
    \label{fig:cas_ag2_sim}
\end{figure}

Next, we wanted \texttt{agent 1} to teleport its state to \texttt{agent 2}. To do this communication, we first implemented the circuit of \texttt{agent 1} from qubits $q_{0}$ through $q_{4}$, followed by the implementation of the teleportation protocol \cite{Bennett1993, Djordjevic2022, Hamdoun2020} on qubits $q_3$, $q_5$, and $q_9$. Then, the initial state of \texttt{agent 2}'s melodic qubit became the state of \texttt{agent 1}'s melodic qubit. After this stage, \texttt{agent 2}'s circuit was implemented between qubits $q_6$ through $q_{10}$. In Figs. \ref{fig:cas_ag1_sim} and \ref{fig:cas_ag2_sim}, we see the simulated results of a two-agent system where \texttt{agent 1} teleported its state to \texttt{agent 2}. 

\medskip
The resulting melodies from \texttt{agent 1} and \texttt{agent 2} in the simulator case can be compared using a melodic correlation measure proposed by \cite{Kim2024}. This calculates the similarity between the contours of the melodic phrases. As the similarity between the phrases increases, the melodic correlation value would also increase, with identical melodies having a value of equal to 1. As the similarity between the phrases decreases, the melodic correlation values would also decrease, with melodies with very different contours being very close to valuing 0. Melodic correlation can be expressed mathematically as shown in Eq. \ref{eq:mel_corr}, where $\Delta$ represents the number of note intervals within the melodies being compared, while $A^{(1)}$ and $A^{(2)}$ represent the notes present in the phrases of \texttt{agent 1} and \texttt{agent 2} respectively  \cite{Nerness2025}. Also note that the addition of $0.1$ in both the numerator and denominator in calculating the ratios of intervals between the agents prevents a zero occurring in the denominator in the case of non-intervallic change. In the simulator results shown above we have pitch sets of $A^{(1)} = \bigl\{F4, G4, A4, F4, F4, G4, C5, D5\bigr\}$ and $A^{(2)} = \bigl\{G4, F4, F4, C5, F4, F4, F4, F4 \bigr\}$. Calculating out the melodic correlation of these two phrases, we obtained a correlation equal to $0.256$. Since $mcorr < 0.5$ we can say that the melodic contours of these two phrases have a weak correlation. 

\begin{equation}
	\normalsize
	mcorr = \frac{1}{\Delta}  \sum_{i=1}^{\Delta}{min\left(\frac{\abs{A^{(1)}_{i} - A^{(1)}_{i-1}}+0.1}{\abs{A^{(2)}_{i} - A^{(2)}_{i-1}}+0.1}, \frac{\abs{A^{(2)}_{i} - A^{(2)}_{i-1}}+0.1}{\abs{A^{(1)}_{i} - A^{(1)}_{i-1}}+0.1} \right)}
\label{eq:mel_corr}
\end{equation}

In addition to comparing the melodic contours, we also considered the distance between the pitch-class sets of these generated melodies. To calculate this, we employed the Hamming distance \cite{Andronikos2026}, expressed as:

\begin{equation}
	\normalsize
	H(A^{(1)}, A^{(2)}) = \sum_{i=1}^{P}{A_{i}^{(1)} \oplus A_{i}^{(2)} }
\label{eq:hdist}
\end{equation}

In Eq. \ref{eq:hdist} the Hamming distance $H(A^{(1)}, A^{(2)})$ between \texttt{agent 1} and \texttt{agent 2} is a function of two binary strings. Each string represents the set of unique pitches within our key centre for a particular agent.  If a specific note represented in the binary string is denoted as a 1, it means that note in the key centre exists in the pitch-class set of the respective agent. If a note in the string is denoted with a 0, then it means that the note is absent from the pitch-class set of the agent. Once binary strings for \texttt{agent 1} and \texttt{agent 2} are constructed, we compare each bit by applying a controlled-\textbf{X} (CNOT) operation between the two bits. If the result is equal to 0, then that means the bits were identical, and if it is equal to 1, that means they were different. As shown in equation \ref{eq:hdist}, we then add up all the results from this comparison to get our Hamming distance.

\medskip

To apply the above to the simulator results, for this demonstration, we first had to create the binary string identifying which notes in our key centre of F major were included in each agent's respective pitch-class sets. Comparing \texttt{agent 1}'s pitch set $\{F, G, A, C, D\}$ with the F major scale $\{F, G, A, Bb, C, D, E\}$, we obtained the binary string of $1110110$. With \texttt{agent 2}'s pitch set, we got the binary string of $1100100$. As a result, plugging these into the Hamming distance equation, we calculated the distance between the agents' pitch sets:  $H(A^{(1)}, A^{(2)}) = 2$. 

\begin{figure}[htbp]
    \centering
    \subfigure[Melodic and rhythmic qubit signals of  \texttt{agent 1}.]{
         \includegraphics[scale=0.5]{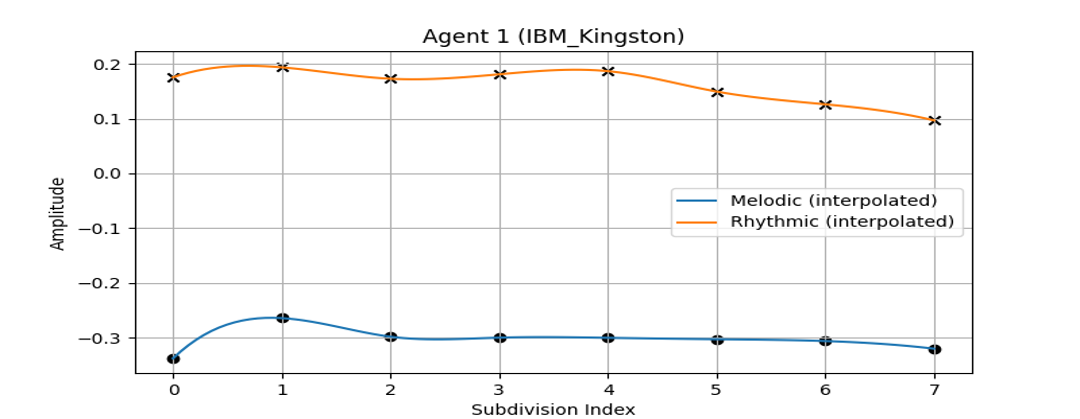}
        \label{fig:subim5}
    }
    \subfigure[Rendered musical notation of \texttt{agent 1}.]{
        \includegraphics[scale=0.75]{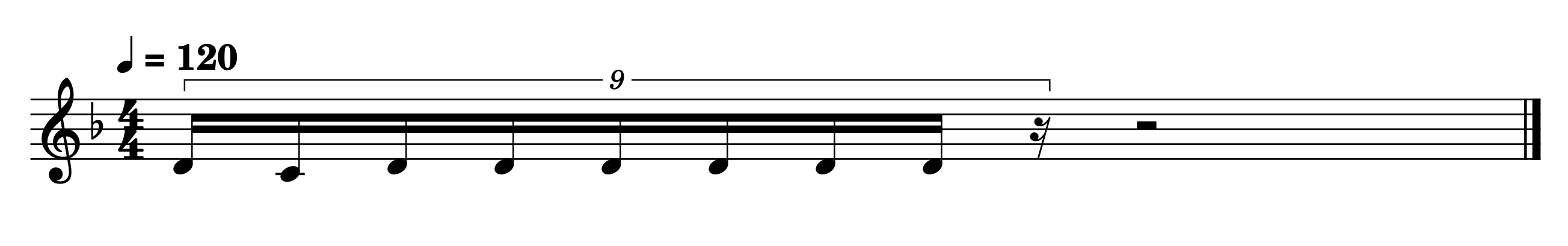}
        \label{fig:subim6}
    }
    
    \caption{Results for \texttt{agent 1} executed on the IBM quantum hardware, ibm\_kingston.}
    \label{fig:cas_ag1_real}
\end{figure}

\begin{figure}[htbp]
    \centering
    \subfigure[Melodic and rhythmic qubit signals of  \texttt{agent 2}.]{
         \includegraphics[scale=0.5]{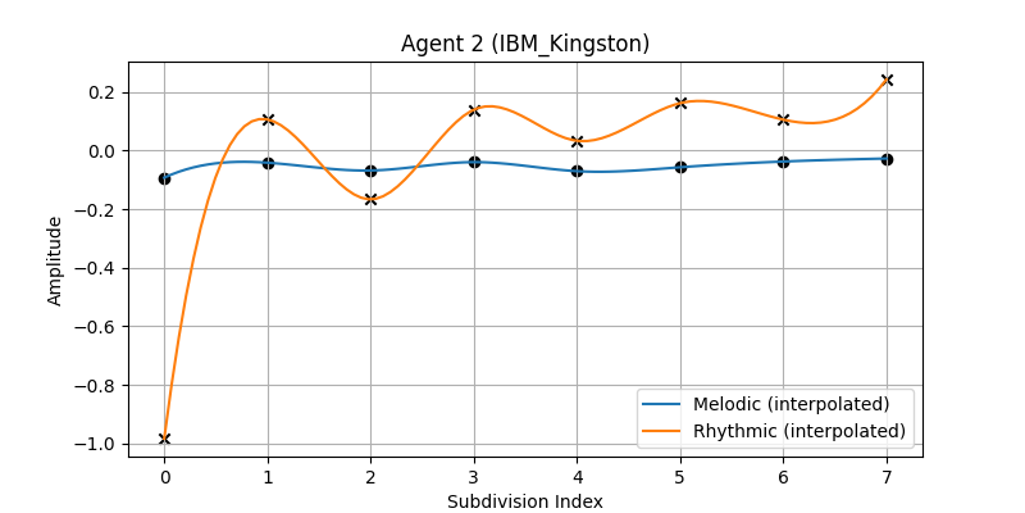}
        \label{fig:subim7}
    }
    \subfigure[Rendered musical notation of \texttt{agent 2}.]{
         \includegraphics[scale=0.75]{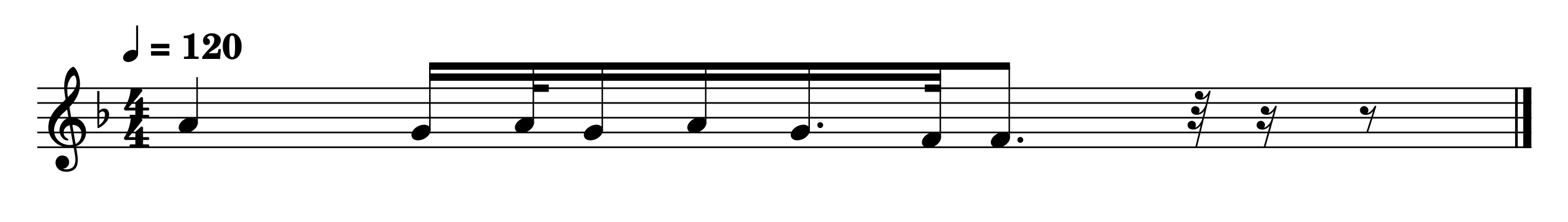}
        \label{fig:subim8}
    }
    \caption{Results for \texttt{agent 2} executed on the IBM quantum hardware, ibm\_kingston.}
    \label{fig:cas_ag2_real}
\end{figure}

Figures \ref{fig:cas_ag1_real} and \ref{fig:cas_ag2_real} show the results of this configuration when run on an IBM quantum hardware \texttt{ibm\_kingston}. The melodic correlation between the resulting melodic phrase was 0.456, showing that for this particular example, that the melodies generated from \texttt{agent 1} and \texttt{agent 2} on the real quantum hardware were more correlated by a factor of 1.78 than the simulator. This may be due to the fact that the addition of the hardware noise may have caused the two agents' melodic states to converge closer together than they did in the agents in the simulator case. However, the Hamming distance for the real hardware case was calculated to be $H(A^{(1)}, A^{(2)}) = 5$, which was greater than the Hamming distance for the results of the simulator case. Even though in the hardware case an increase in melodic correlation was observed, we also saw an increase in the distance between the pitch sets of the agents.

\subsection{Three Agents Example}
\label{sec:cas_three}

For this example, we scaled the system up to three agents arranged in a unidirectional cascade as shown in Fig. \ref{fig:Multiagent_Circuit}. In this configuration, quantum states originating from one agent influenced two downstream agents, either \textit{directly} or \textit{indirectly}, through successive teleportation-based interaction steps. 

\medskip

Direct influence occurs when one agent teleports their state to another agent. An example of this is the interaction between \texttt{agent 1} and \texttt{agent 2} in the previous section. Indirect influence occurs when an agent teleports its state through one or more agents to reach the final receiving agent. For example, in the example below, \texttt{agent 1}'s state was teleported and transformed through \texttt{agent 2}, then finally teleported to \texttt{agent 3}. As a result, \texttt{agent 1} indirectly influenced \texttt{agent 3}'s final quantum state.

\begin{table}[htbp]
    \centering
    \begin{tabular}{|c|c|c|}
        \hline
        Agent Pair & Melodic Correlation & Hamming Distance \\
        \hline
        \texttt{Agent 1}, \texttt{Agent 2} & 0.135 & 2 \\
        \texttt{Agent 1}, \texttt{Agent 3} & 0.109 & 2 \\
        \texttt{Agent 2}, \texttt{Agent 3} & 0.717 & 0\\
        \hline
    \end{tabular}
   \caption{Melodic correlation and pitch set Hamming distance between pairs of agents, with Qiskit's Aer simulator.}
    \label{tab:3c_results_sim}
\end{table}

\begin{figure}[htbp]
    \centering
    \subfigure[Melodic and rhythmic qubit signals of  \texttt{agent 1}.]{
	\includegraphics[scale=0.5]{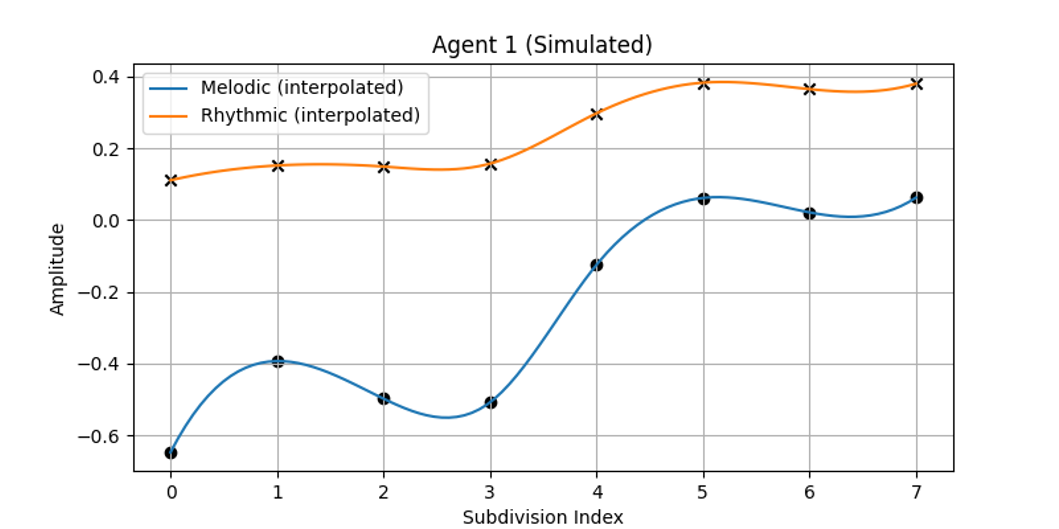}
    \label{fig:3c_a1_sim}
    }
   \subfigure[Rendered musical notation of \texttt{agent 1}.]{
        \includegraphics[scale=0.75]{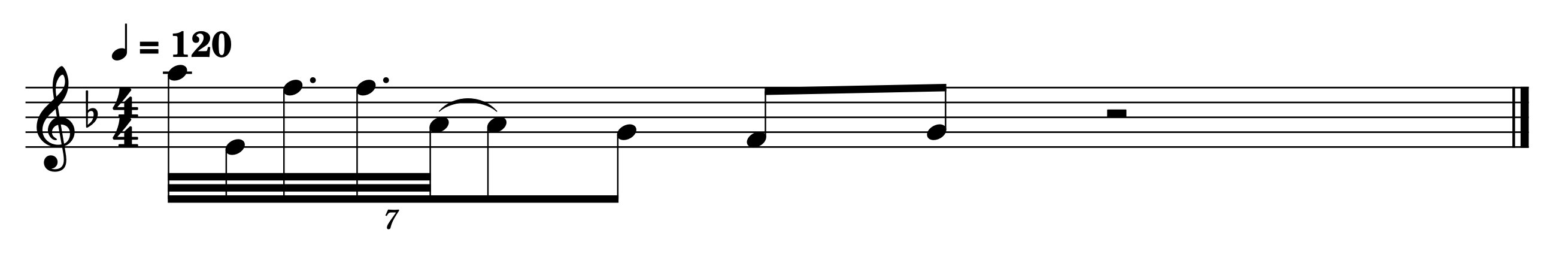}
         \label{fig:3c_a1_music_sim}
    }
    \caption{Results for \texttt{agent 1} in a 3-agent cascading scenario executed on Qiskit's Aer simulator.}
    \label{fig:3c_ag1_sim}
\end{figure}

\begin{figure}[htbp]
    \centering
    \subfigure[Melodic and rhythmic qubit signals of  \texttt{agent 2}.]{
	\includegraphics[scale=0.5]{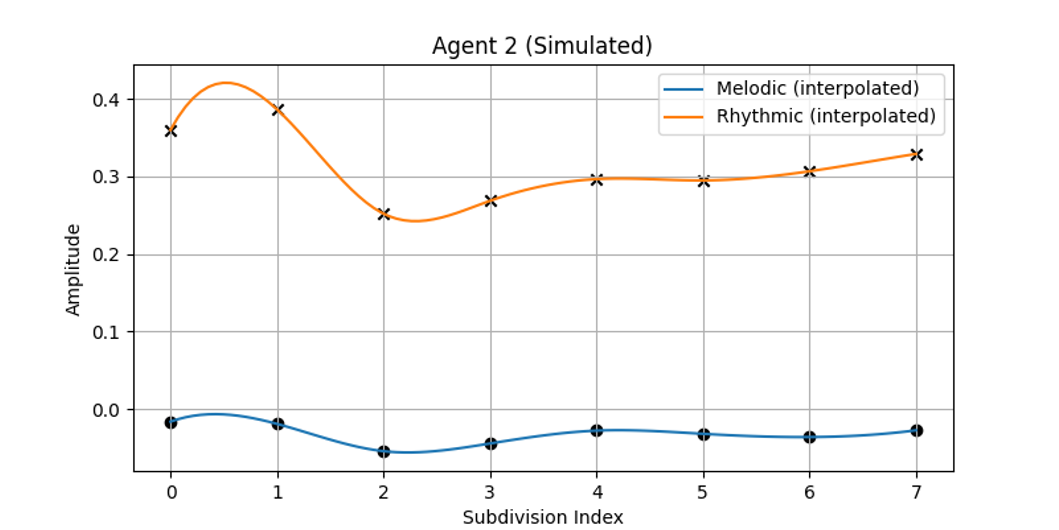}
    \label{fig:3c_a2_sim}
    }
   \subfigure[Rendered musical notation of \texttt{agent 2}.]{
        \includegraphics[scale=0.75]{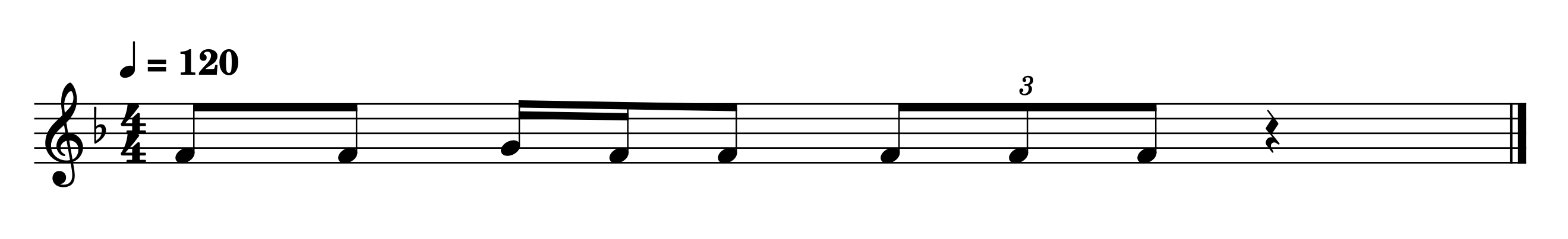}
         \label{fig:3c_a2_music_sim}
    }
    \caption{Results for \texttt{agent 2} in a 3-agent cascading scenario executed on Qiskit's Aer simulator.}
    \label{fig:3c_ag2_sim}
\end{figure}

\begin{figure}[htbp]
    \centering
    \subfigure[Melodic and rhythmic qubit signals of  \texttt{agent 3}.)]{
	\includegraphics[scale=0.5]{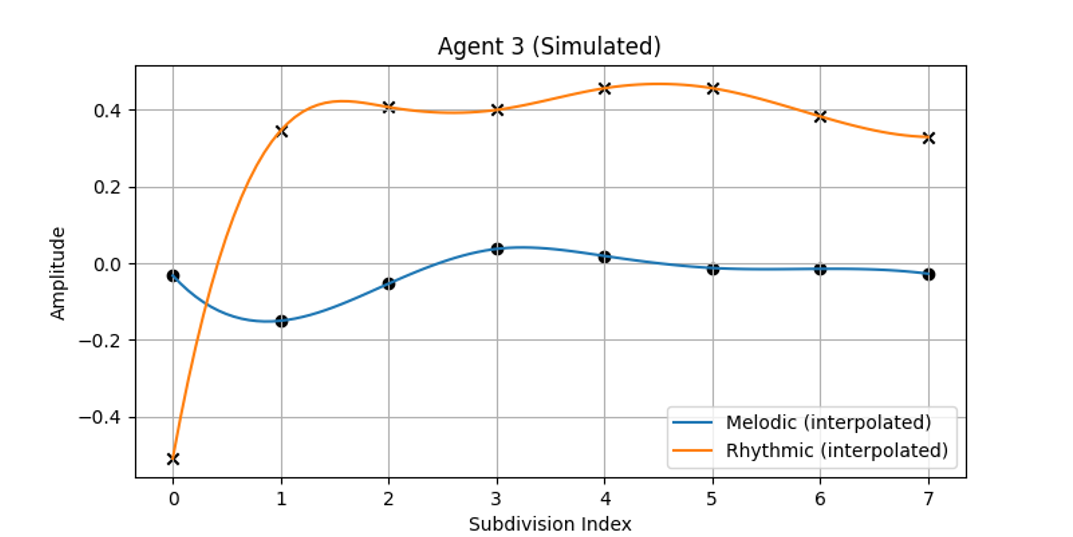}
    \label{fig:3c_a1_sim}
    }
   \subfigure[Rendered musical notation of \texttt{agent 3}.]{
       \includegraphics[scale=0.75]{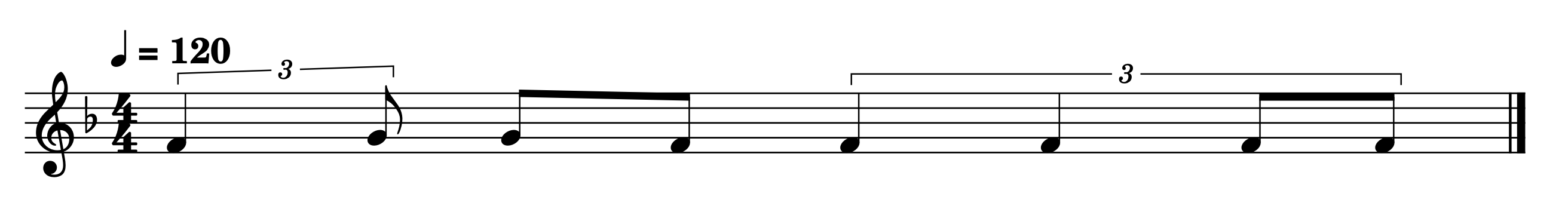}
         \label{fig:3c_a2_sim}
    }
    \caption{Results for \texttt{agent 3} in a 3-agent cascading scenario executed on Qiskit's Aer simulator.}
    \label{fig:3c_ag3_sim}
\end{figure}

The results from running the system on the Aer simulator are shown in Figs. \ref{fig:3c_ag1_sim}, \ref{fig:3c_ag2_sim}, \ref{fig:3c_ag3_sim}. The melodic correlation between \texttt{agent 1} and \texttt{agent 2}, \texttt{agent 1} and \texttt{agent 3}, and finally \texttt{agent 2} and \texttt{agent 3} were calculated along with the Hamming distance for each agent's pitch sets. These results are shown in Table \ref{tab:3c_results_sim}. The melodic contours that are most correlated between these three agents are \texttt{agent 2} and \texttt{agent 3} with a melodic correlation equal to 0.717. In addition, the Hamming distance between  \texttt{agent 2} and  \texttt{agent 3} is $H(A^{(2)}, A^{(3)}) = 0$, meaning that these agents shared a pitch set. The melodic correlations between the other agent pairs were significantly weaker. For  \texttt{agent 1} and  \texttt{agent 2}, the melodic correlation was equal to 0.135, while for  \texttt{agent 1} and  \texttt{agent 3}, the value was equal to 0.109.  The Hamming distance also reflected those weak correlations. Both distances between the pitch sets of  \texttt{agent 1} and \texttt{agent 2}, and between \texttt{agent 1} and  \texttt{agent 3}, were equal to 2. Conversely, the Hamming distance between  \texttt{agent 2} and  \texttt{agent 3} was equal to 0, showing that these agents shared the same pitch set. Thus, it is clear that \texttt{agent 3} was primarily driven by the direct influence of \texttt{agent 2}.

\begin{figure}[htbp]
    \centering
    \subfigure[Agent 1 melodic and rhythmic qubit signals.]{
	 \includegraphics[scale=0.6]{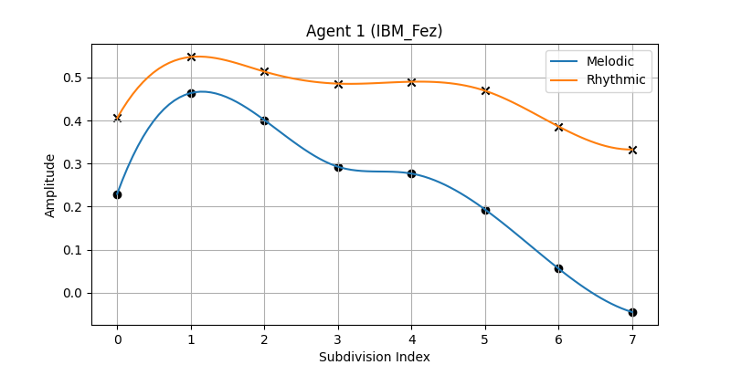}
    \label{fig:3c_a1_fez}
    }
   \subfigure[Agent 1 rendered musical score.]{
        \includegraphics[scale=0.75]{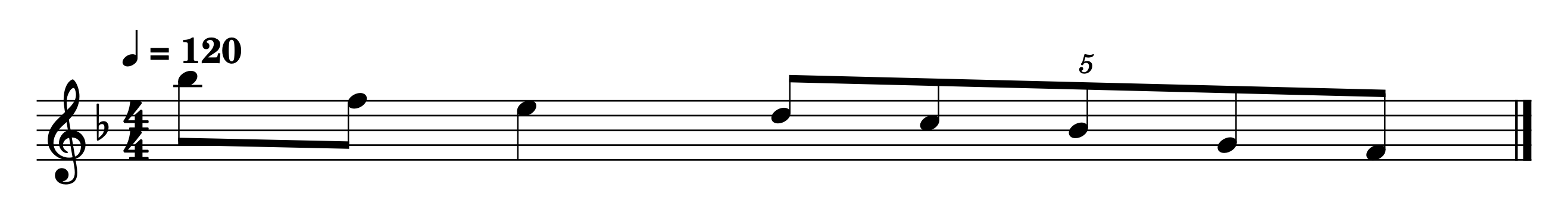}
         \label{fig:3c_a1_music_fez}
    }
    \caption{Results for Agent 1 in a 3-agent cascading scenario executed on IBM quantum hardware, ibm\_fez.}
    \label{fig:cascade-ibm-fez-A}
\end{figure}

\begin{figure}[htbp]
    \centering
    
    \subfigure[Agent 2 melodic and rhythmic qubit signals.]{
    \includegraphics[scale=0.6]{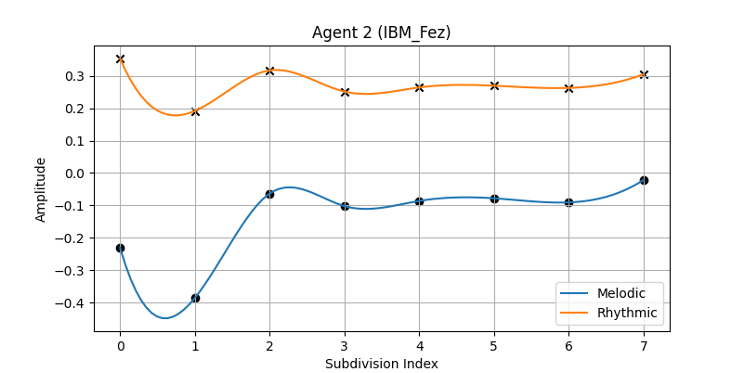}	 
    \label{fig:3c_a2_fez}
    }
   \subfigure[Agent 2 rendered musical score.]{
      \includegraphics[scale=0.75]{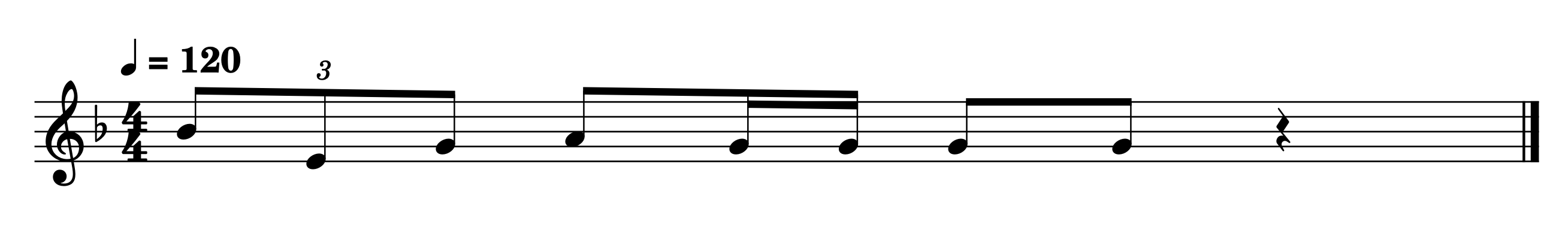}
         \label{fig:3c_a2_music_fez}
    }
    \caption{Results for Agent 2 in a 3-agent cascading scenario executed on IBM quantum hardware, ibm\_fez.}
    \label{fig:cascade-ibm-fez-B}
\end{figure}

\begin{figure}[htbp]
    \centering
    \subfigure[Agent 3 melodic and rhythmic qubit signals.]{
	\includegraphics[scale=0.6]{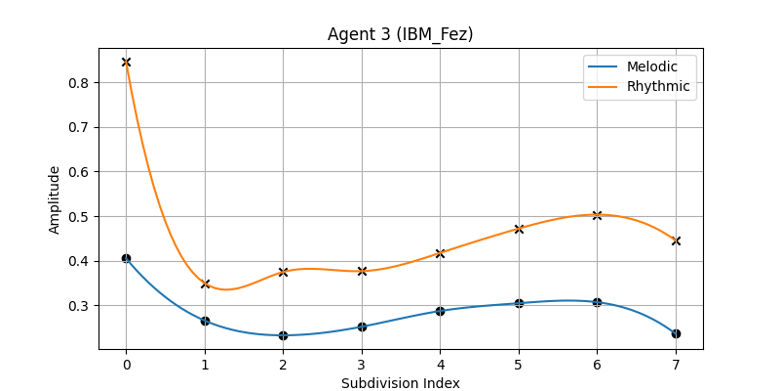}
    \label{fig:3c_a3_fez}
    }
   \subfigure[Agent 3 rendered musical score.]{
      \includegraphics[scale=0.75]{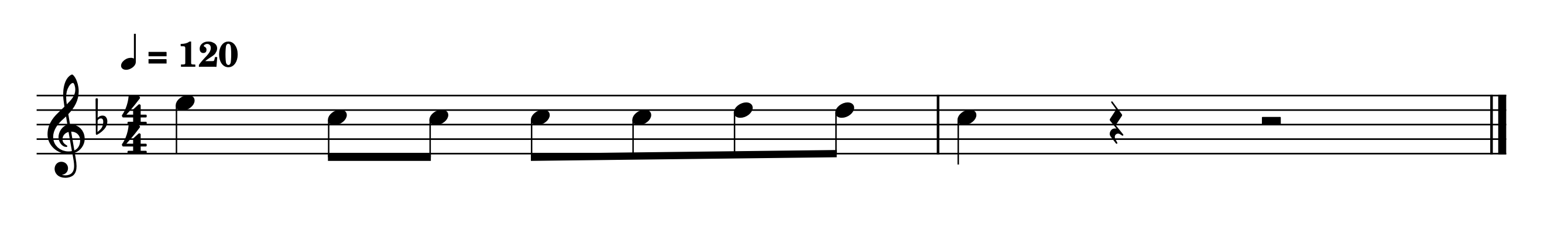}
         \label{fig:3c_a3_fez}
    }
    \caption{Results for Agent 3 in a 3-agent cascading scenario executed on IBM quantum hardware, ibm\_fez.}
   \label{fig:cascade-ibm-fez-C}
\end{figure}

Next,  Figs. \ref{fig:cascade-ibm-fez-A}, \ref{fig:cascade-ibm-fez-B}, \ref{fig:cascade-ibm-fez-C} show the results for running the quantum circuit on the IBM quantum hardware, ibm\_fez. In this case, we observed that  \texttt{agent 1} and  \texttt{agent 2} had the highest melodic correlation with a value equal to 0.497. \texttt{Agent 1} and  \texttt{agent 3} displayed the second-highest correlation with a slightly lower value for melodic correlation, equal to 0.432. Then, \texttt{agent 2} and  \texttt{agent 3} had the smallest melodic correlation, with a value equal to 0.290 (Table \ref{tab:3c_results_real}). 

\begin{table}[htbp]
    \centering
    \begin{tabular}{|c|c|c|}
        \hline
        Agent Pair & Melodic Correlation & Hamming Distance \\
        \hline
        \texttt{Agent 1}, \texttt{Agent 2} & 0.497 & 4\\
        \texttt{Agent 1}, \texttt{Agent 3} & 0.432 & 3\\
        \texttt{Agent 2}, \texttt{Agent 3} & 0.290 & 5\\
        \hline
    \end{tabular}
   \caption{Melodic correlation and pitch set Hamming distance between pairs of agents, with the ibm\_fez processor.}
    \label{tab:3c_results_real}
\end{table}

Table \ref{tab:3c_results_real} shows that \texttt{agent 1} had direct influence on \texttt{agent 2} and indirect influence on  \texttt{agent 3}. It also shows that  \texttt{agent 2}'s direct influence over \texttt{agent 3}'s state was lower than \texttt{agent 1}'s indirect influence, which is contrary to what we observed in the simulator case. The Hamming distance also reflected this behaviour, with the distance between \texttt{agent 2} and \texttt{agent 3}'s pitch sets being the furthest, with a value of 5. The discrepancy between using a quantum computing simulator and using real hardware was somewhat expected. Nevertheless, to move this research forward, we ought to analyse this phenomenon further to better understand the cascading behaviour.

\section{Analysis of Interpretative Distance}
\label{sec:monte_carlo}

This section shows preliminary results from experimenting with interpretive distance tuning parameters for quantum agent interactions embedded within the quantum teleportation protocol. The demonstrations presented above (Section \ref{sec:multi-agents-demo}) showed that interaction between the quantum agents can generate materials ranging from closely aligned imitations to substantial divergences. Here, we formalise this behaviour by analysing agent interpretation. That is, we analysed the extent to which a receiving agent preserves, transforms, or distorts the signal teleported from another agent. Rather than treating such variation as unpredictable artefacts, we examined how interpretative distance might be analysed and controlled within the proposed framework. 

\medskip

For the interpretive distance analysis, we propose a variation to the quantum teleportation protocol. The teleportation protocol itself offers a locus for shaping such interpretation. 

\medskip

\begin{figure}[htbp]
\begin{center}\vspace{0.3cm}
    \begin{tikzpicture}
        \node[scale=1.0] 
        {
            \begin{quantikz}
                	\lstick{$q_0$} & \ket{\Psi}_1 & \gate[style={fill=yellow!30}]{\textbf{U}(\theta, \phi, \lambda)} & \hphantomgate{} & \ctrl{1} & \gate{\textbf{H}}  & \meter{}   \\
                	\lstick{$q_1$} & \ket{0} & \gate{\textbf{H}} & \ctrl{1} & \gate{\textbf{X}}  & \meter{}   \\
                	\lstick{$q_2$} & \ket{0} &  \hphantomgate{} & \gate{\textbf{X}} & \hphantomgate{} & \gate{\textbf{RX($\theta$)}} \wire[u][1]{c} & \gate{\textbf{RZ($\theta$)}} \wire[u][2 ]{c} &  \ket{\Psi}_2 & \meter{} \		
            \end{quantikz} 
        };
    \end{tikzpicture}
\end{center}
\caption{A variation of the standard quantum teleportation protocol for the interpretive distance analysis.}
\label{fig:Teleportation_Explanation_tuneable}
\end{figure}
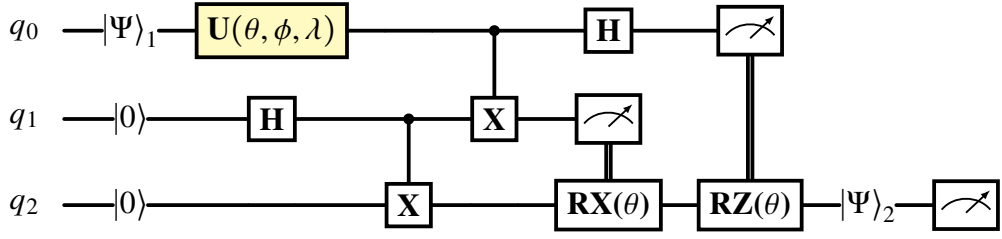

In the standard quantum teleportation protocol circuit (Fig. \ref{fig:Teleportation_Explanation}), Pauli correction gates (\textbf{X} and \textbf{Z}) are applied conditionally to recover the original state, as outlined in section \ref{sec:teleport}. For the interpretive distance analysis, we generalise this correction stage by replacing standard Pauli gates with parameterised rotation gates, \textbf{RX} and \textbf{RZ}, as shown in Fig. \ref{fig:Teleportation_Explanation_tuneable}. 

To explore how agent interpretation varies as a function of \textbf{U} gate parameters, we conducted an analysis over randomly sampled rotations. For each trial, agents were configured with parameterised single-qubit \textbf{U} gates of the form $U(\theta, \phi, \lambda)$, where the parameters correspond to rotations in computational and phase bases, as described in Sections \ref{sec:single_agent_circuit} and \ref{sec:multi-agents-demo}. Each parameter set  $p = (\theta, \phi, \lambda)$ was sampled uniformly from a predefined range and grouped into a vector of trials (Eq. \ref{eq:fidelity}).

\begin{equation}
    \Theta = (p_1, p_2, p_3, ... p_k)
    \label{eq:fidelity}
\end{equation}

For each trial, the teleportation protocol was executed between two agents, and the resulting density matrices of the melodic and rhythmic qubits were reconstructed from measurement statistics. Fidelity was then computed pairwise between sender and receiver states. Fidelity here provides a scalar measure of overlap between two quantum states and is commonly used to evaluate the quality of state transmission in quantum communication systems. In this context, however, fidelity was reinterpreted as a measure of musical understanding between agents. Specifically, we used the Uhlmann-Jozsa Fidelity measure as presented in \cite{Liang2019}.

\medskip

\begin{figure}[htbp]
    \centering
    \subfigure[Normal distribution for state fidelities.]{
         \includegraphics[width=0.85\linewidth]{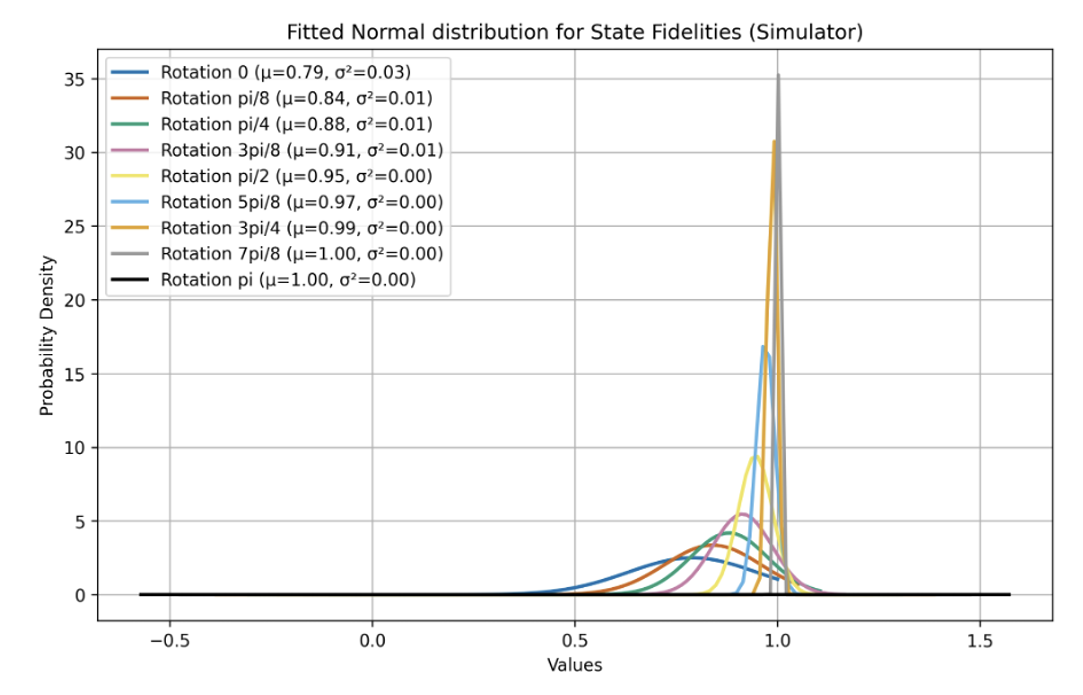}
       \label{fig:Fidelity-1} 
    }
    \subfigure[State fidelity results, between \texttt{agent 1} and \texttt{agent 2}.]{
      \centering
      \includegraphics[width=0.75\linewidth]{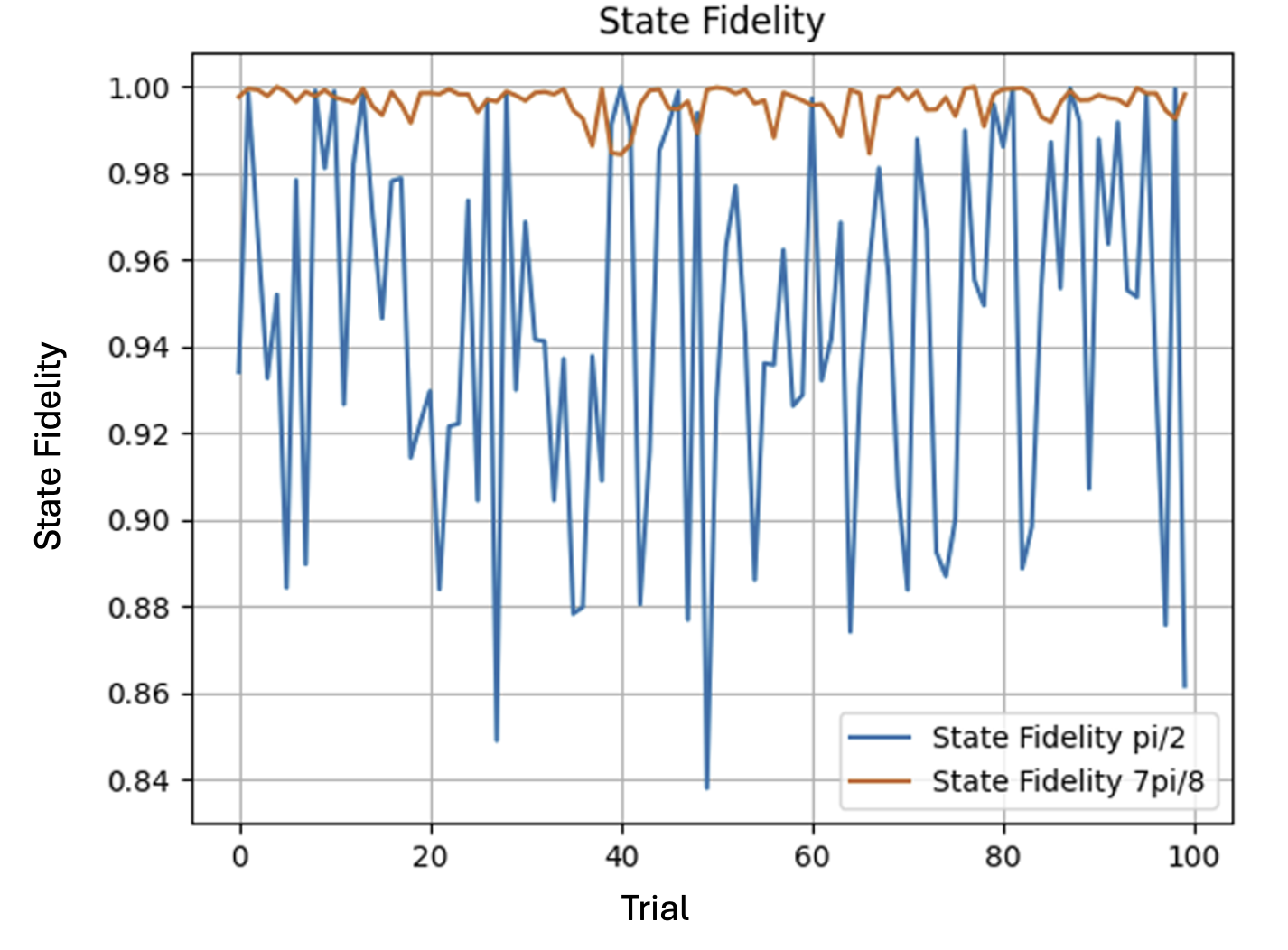}
        \label{fig:Fidelity-2} 
    }
    \caption{Results of interpretive distance analysis.}
    \label{fig:Fidelity}
\end{figure}

Given two quantum states represented by density matrices $\rho_1$ and $\rho_2$, the state fidelity is defined as shown in Eq. \ref{eq:voyager}. Figure \ref{fig:Fidelity} (a) aggregates 100 trials on Qiskit's Aer simulator. Figure \ref{fig:Fidelity-1} summarises the resulting fidelity distributions for representative parameter regimes.

\begin{equation}
    F(\rho_1, \rho_2) = (\texttt{Tr} [ \sqrt{ \sqrt{\rho_1} \rho_2 \sqrt{\rho_1}}] )^2
    \label{eq:voyager}
\end{equation}

For each trial, the teleportation protocol was executed between two agents, and the resulting density matrices of the melodic and rhythmic qubits were reconstructed from measurement statistics. Fidelity was then computed pairwise between sender and receiver states. The results show systematic trends rather than random variation. On the simulator, increasing the rotation angle $\theta$ towards $\pi$ leads to higher mean fidelity and reduced variance, indicating more stable transmission. Conversely, smaller rotations, near zero, result in lower mean fidelity and greater dispersion, corresponding to increased interpretative variability. These findings support the interpretation of teleportation-based interactions not as binary processes (i.e., successful versus failed), but as a continuous space of interpretative relationships shaped by gate parameters and hardware characteristics. 

The procedure above allows for interpretative behaviour to be tuned deliberately, rather than emerging solely from hardware-induced noise. From a musical perspective, it enables agents to adopt distinct \enquote{listening styles}, analogous to performers who respond with varying degrees of precision, exaggeration, or abstraction. We refer to this configurable correction stage as an \textit{interpreter module}. For example, replacing an \textbf{X} gate correction with an $\textbf{RX}(\theta)$ gate and varying $\theta$ from $\pi$ to $0$ produces a gradual decrease in mean fidelity and an increase in variance across trials. It is important to note that the examples presented in sections \ref{sec:single_agent_circuit} and \ref{sec:multi-agents-demo} use a rotation angle of $\theta = \pi$. Thus, all instances of melodic convergence and divergence between agents were observed using the baseline teleportation protocol.

\medskip
Conducting such an experiment on the IBM hardware would have required resources that were beyond our affordability at the time. However, we assume that the noise associated with the hardware devices will increase the variance of each of the rotation distributions. For future research, we ought to understand the role of hardware noise, especially when implementing multi-agent systems on a quantum network. 

\section{Concluding Discussion}
\label{sec:conclusion}

We presented a framework for interactive music based on quantum agents that communicate through quantum teleportation. Rather than treating quantum computation as a black-box generative mechanism, the framework emphasises agent-level behaviours, directed influence, and probabilistic interpretation as expressive musical resources.

\medskip

Our quantum agents were not conceived as deterministic music generators. Rather, they were developed to behave like semi-autonomous musical performers whose interactions resemble the conversational, imitative, and occasionally distorted exchanges characteristic of free Jazz improvisation. Here, we proposed the notion of \textit{quantum whisper} as a metaphor for agent communication in which divergence and ambiguity are tolerated. 

\medskip

In Section \ref{sec:single_agent_circuit}, we introduced the foundational behaviour of a single agent. By combining the SQPAM representation scheme with the PKBSE technique, individual agents generated musical material that is structured yet inherently variable. Repeated executions of the same circuit yielded related but non-identical outputs, reflecting the probabilistic nature of quantum measurement and the influence of hardware-induced noise. This behaviour aligns with established practices in algorithmic and improvised music, where variation within constraint is a central compositional principle. It also demonstrates that meaningful musical structures can be extracted from compact quantum representations using a small number of qubits, even under NISQ-era limitations.

\medskip

Then, in the two-agent configuration introduced in Section \ref{sec:cas_two}, quantum teleportation was presented as a mechanism for directed musical influence. Unlike entanglement-based coupling, teleportation allowed one agent's internal state to shape another's behaviour without enforcing synchronisation or identity. Musical materials were consequently transformed rather than duplicated. Notably, the temporal misalignment between agents' time registers caused teleported states to manifest at different sequence positions. This contributed meaningfully to musical divergence. This phenomenon warrants deeper investigation in future work as a controllable expressive parameter rather than a limitation.

\medskip

Extending the system to three agents in Section \ref{sec:cas_three} revealed how cascading teleportation-based interactions can produce emergent ensemble behaviour. The results from both the simulator and quantum hardware demonstrated that downstream agents embody composite influences: in the simulator, 
\texttt{agent 3} was primarily driven by its immediate predecessor, \texttt{agent 2}, while on quantum hardware, \texttt{agent 1}'s indirect influence appeared to dominate. This highlighted how NISQ noise itself reshapes the character of agent interaction in musically significant ways. Rather than treating these divergences as errors, we proposed to consider them as a continuum of interpretative distance.

\medskip

This reframing was formalised in Section \ref{sec:monte_carlo}.  The standard teleportation correction stage was generalised into a tunable interpreter module. By replacing Pauli correction gates with parameterised rotation gates, we demonstrated that interpretative behaviour can be systematically shaped. Interpretative behaviour is the degree to which a receiving agent preserves or transforms a teleported state. Increasing the rotation angle $\theta$ of a parameterised rotation gate (e.g., $\textbf{RX}(\theta)$ towards $\pi$ produced higher mean fidelity and reduced variance. Conversely, smaller rotations increased interpretative variability. This supports a model of agent interaction as a continuous space of musical understanding, rather than a binary success or failure of state transmission.

\medskip

Taken together, these results position quantum teleportation as a genuinely promising interaction mechanism for agent-based computer music. Looking forward, the most immediate priorities are to conduct a deeper analysis of temporal misalignment effects in multi-agent systems and the development of performer-facing controls over teleportation and interpretation parameters.  Most ambitiously, as quantum hardware and networking mature, the vision of geographically distributed agents connected through the Quantum Internet offers the prospect of musical performance practices that are fundamentally incommensurable with classical networks, not only in terms of higher speed or lower latency, but qualitatively different in terms of the actual information that is exchanged.

\medskip

With this work, we established a creative paradigm that is genuinely native to quantum computation. Previous algorithmic music systems, however sophisticated, ultimately operated on deterministic or pseudo-random processes.  Our system's outputs are something qualitatively different: they emerge from quantum measurement, superposition, and the probabilistic nature of state transfer. The \textit{quantum whisper} aesthetic embraces noise, decoherence, and interpretative divergence as expressive resources rather than engineering problems. In the longer term, if geographically distributed agents were connected through the Quantum Internet, the nature of musical communication between them would be fundamentally non-local, carrying information in a form that classical networks cannot transmit. 

\medskip

From the perspective of quantum computing and communications, the work contributes something less obvious but equally valuable: a perceptually rich, human-interpretable testbed for studying the qualitative character of imperfect quantum state transfer. Most quantum communication research evaluates teleportation against binary or scalar fidelity metrics, which implicitly assume that deviation from the intended state is simply failure. Our reframing of fidelity as a continuum of interpretative distance might be a more honest account of what happens on NISQ hardware. Music turns out to be a surprisingly adequate medium for making that continuum legible. A listener can perceive and evaluate the relationship between two melodic phrases, whether they are closely imitative, loosely related, or entirely divergent, in ways that a single fidelity score does not capture. This has concrete implications for quantum network research. 

\medskip

The three-agent cascading experiments directly probe how quantum states accumulate transformations across multiple hops, which is a central challenge in Quantum Internet design. The interpreter module, which replaces standard Pauli correction gates with parameterised quantum gates, has a natural analogue in quantum error correction and channel capacity research, where the trade-off between faithful transmission and tolerated transformation is a fundamental design question. As quantum networks scale toward real-world deployment, evaluation frameworks will need to go beyond raw technical metrics. A music system like the one we introduced in this paper may offer a productive bridge between the engineering requirements of quantum communication and the human experience of the information being transmitted.

\section{Acknowledgements}

This work was funded by the University of Plymouth, UK.

\medskip

The authors gratefully acknowledge James (Jim) Weaver of IBM Quantum for his careful review of this manuscript. His thoughtful comments, constructive suggestions, and technical insights have contributed significantly to improving the clarity, accuracy, and overall quality of this work. His willingness to share his time and expertise is sincerely appreciated. Please note that the views expressed in this paper, as well as any remaining errors or omissions, are solely those of the authors and should not be interpreted as reflecting the views of IBM Quantum.



\end{document}